\documentclass{article}
\usepackage{amsmath,amsthm,amssymb,latexsym,cite,finalA}

\numberwithin{equation}{section}

\newcommand{\sauthor}{Oliynyk and K\"unzle}
\newcommand{\stitle}{Spherical EYM fields with compact gauge groups}

\input eymA.def

\begin{document}

\title{Local existence proofs for the boundary value problem for static spherically symmetric Einstein-Yang-Mills fields with compact gauge groups
\thanks{Supported in part by NSERC grant A8059.}
}
\author{
Todd A. Oliynyk \thanks{toliynyk@ualberta.ca}
\and
H.P. K\"{u}nzle \thanks{hp.kunzle@ualberta.ca}\\
Department of Mathematical Sciences, University of Alberta\\
Edmonton, Canada T6G 2G1}

\date{}
\maketitle
\begin{abstract}
  We prove local existence and uniqueness of static spherically symmetric
  solutions of the Einstein-Yang-Mills equations for an arbitrary compact
  semisimple gauge group in the so-called regular case. By this we mean the
  equations obtained when the rotation group acts on the principal bundle on
  which the Yang-Mills connection takes its values in a particularly simple
  way (the only one ever considered in the literature). The boundary value
  problem that results for possible asymptotically flat soliton or black hole
  solutions is very singular and just establishing that local power series
  solutions exist at the center and asymptotic solutions at infinity amounts
  to a nontrivial algebraic problem. We discuss the possible field equations
  obtained for different group actions and solve the algebraic problem on
  how the local solutions depend on initial data at the center and at
  infinity.
\end{abstract}

\sect{intro}{Introduction}

Over the last dozen years much has been learned about the classical
interaction of \YM\ fields with the gravitational field of Einstein's general
relativity. Most investigations have concentrated on \YM-fields with the gauge
group $SU(2)$ starting with Bartnik and Mckinnon's \cite{k4752} discovery of
globally regular and asymptotically flat numerical solutions.  Their global
existence was analytically proved \cite{k5061,k5428} and many further
properties like stability of these particle-like or soliton solutions and the
corresponding black hole solutions were investigated numerically as well as
analytically. Moreover, many different matter fields can be minimally coupled
to the gravitational and \YM\ fields, and corresponding spherically symmetric
solutions have been, mostly numerically, but sometimes also analytically
studied. We refer for the (hundreds of) references to the review article
\cite{k6373}.

Some similar phenomena were found for special models with gauge groups $SU(n)$
for $n>2$ \cite{k5187,hka26,k6174,k6485}, and the general static spherically
symmetric equations for general compact gauge groups were derived already
quite early \cite{k5109,k5281}.

For larger gauge groups than $SU(2)$ the notion of spherical symmetry
is no longer straight forward enough for a simple ansatz to work.
Instead one needs to consider the possible actions of the symmetry group
$SO(3)$ or $SU(2)$ by automorphisms of principal bundles over space-times whose
structure group $G$ is the gauge group of the \YM\ field.  A conjugacy class
of such automorphisms is characterized by a generator $\Lo$ which is an
element of a Cartan subalgebra $\h$ of the complexified Lie algebra $\g$ of
$G$ \cite{k5109,k5281}. Mostly one restricts consideration to fields which are
regular at the center or, for black hole fields, to those for which the
\YM-curvature falls off sufficiently fast at infinity. In \cite{k6217} these
are called these regular models. They also correspond to the ``no magnetic
charge'' case in \cite{k4295}. For these group actions the element $\Lo$ of
$\h$ must be an \emph{\Aonev} or \emph{defining vector} of an
$\sL(2)$-subalgebra of $\g$.

That there is a remarkable variety of possible actions was shown by Bartnik
\cite{k6217} for the case where $G$ is any group with Lie algebra $\su(n)$.
More generally, for arbitrary semisimple Lie algebras these \Aonev{s} were
classified by Mal'cev \cite{k6436} and Dynkin \cite{k4779} and can now also be
obtained more conveniently from the theory of nilpotent orbits \cite{k6494}.

One of these classes of actions of the symmetry group is somewhat
distinguished. It corresponds to a principal \Aonev\ in Dynkin's terminology
and we will call it a \emph{principal action}. To our knowledge almost all
work for larger gauge groups has been done for this case \cite{hka28,k5749,
  k6485,k6299}. For a slightly bigger class of actions, the ``generic'' class
in \cite{k4295} which we will call \emph{regular}, the \Aonev\ lies in the
interior of a fundamental Weyl chamber. Brodbeck and Straumann
\cite{k4295,k5626} proved that all regular static asymptotically flat
solutions are unstable against time dependent perturbations. They were able to
do this without establishing existence or any properties of these solutions.

While it is easy to show that, at least for the regular case, some global
solutions exist, namely those which arise by scaling from some imbedded
$SU(2)$ solutions, only isolated, mostly numerical, results have been obtained
about more general global solutions for the principal $SU(n)$ actions for
$n=3,4,5$ \cite{hka28,k5749,k5955,k6485,k6299}.

The purpose of this paper is to discuss the classification of all the
regular actions of $SU(2)$ by automorphisms of $G$-principal bundles over
spherically symmetric static space-times and to analyze the resulting \EYM\ 
field equations to the extent of establishing that the singular boundary value
problem obtained for the globally regular and asymptotically flat solutions
is well defined ``at both ends'', namely at the center or the black hole
horizon and at infinity. 
The local solutions that we obtain near these points 
are actually analytic. Consequently, there exists
convergent powerseries representations for these solutions
at least for small distances from the center, the black hole
horizon and infinity.
Essentially we generalize the results of
\cite{hka28} from the principal action on $SU(n)$-bundles to regular actions
on bundles with (simply connected) semisimple compact structure groups.
Although this represents only a first step in an analysis of possible (non
scaled) global solutions establishing these local existence theorems is
already quite complicated.  It is worthwhile to note, that if
any of the local solutions can be extended to a
global one, then the results of Brodbeck and Straumann \cite{k5626}
apply and show that the solution must be unstable.

For all these regular models it turns out that the Yang-Mills potential can
be chosen (i.e. suitably gauged) to depend only on $\ell$ real-valued
functions of a radial coordinate $r$ where $\ell$ is the rank of the Lie
algebra of the gauge group. In addition, the metric will be given by two more
functions of $r$. These $(\ell+2)$ functions satisfy a nonlinear system of
ordinary differential equations which has singularities at $r=0$, when
$r\ra\infty$, and at the horizon where $r=\rh$, say. We need to analyze these
singularities to determine the ``initial conditions'' for these functions and
the number of free parameters that can be chosen when solving the equations
numerically, for example, by the method of shooting to a meeting point. In
this paper we will only establish what these parameters are, we will not solve
the equations numerically.

There are many models for which the \Aonev\ $\Lo$ is on the
boundary of a Weyl chamber. To our knowledge almost no results have been
obtained for them, but we have reason to believe that some of our methods may
also be useful for these irregular models.

In section \ref{class} we review the description of the class of static
spherically symmetric models and in section \ref{eqs} we show, starting from
the field equations, that the special class of models we call regular can be
reduced to the principal case for imbedded semisimple groups. We discuss the
initial value problems somewhat informally in section \ref{sol} where we
derive the relatively complicated way in which a solution depends on
parameters chosen at the endpoints of the $r$-interval. In section \ref{elem}
we extend some elementary facts that are well known for $SU(2)$-solutions to
general compact $G$. Finally, the major part of this paper consists of the
proofs, divided into section \ref{BFM1} containing algebraic lemmas and the
proof of the local existence theorems for the differential equation system in
section \ref{BFM2}.

\sect{class}{Classes of spherically symmetric Yang-Mills connections}

Since there is no natural action of the symmetry group on the principal bundle
we need to consider all possibilities, i.e. all conjugacy classes of actions
of $SO(3)$, or for simplicity, $SU(2)$ by automorphisms on principal
$G$-bundles $P$ over space-time $M$ which project onto isometries of $M$ with
orbits diffeomorphic to 2-spheres. We assume throughout that $G$ is a compact
semisimple connected and simply connected Lie group.

Then these conjugacy classes are in one-to-one correspondence with integral
elements $\Lo$ of the closed fundamental Weyl chamber $\overline{W(S)}$
belonging to some basis $S$ of the roots of $\g$ for some chosen Cartan
subalgebra $\h$ \cite{k5109,k5281,k6217}. Here $\g=(\g_{0})_{\CO}$ stands for
the complexification of the Lie algebra $\g_{0}$ of the structure group $G$ of
$P$. If $\{\tau_{i}\}$ is a standard basis of the Lie algebra $\su(2)$ such
that $[\tau_{i},\tau_{j}]=\epsilon_{ij}^{\phantom{ij}k}\tau_{k}$ then $\Lo$
may be chosen such that 
\eqn{lam0}{ 
\Lo = 2 i\;\lambda(\tau_{3}) 
} 
where $\lambda$ is the (induced Lie algebra) homomorphism from the isotropy
group $I_{x_{0}}$ of the $SU(2)$-action on $M$ at $x_{0}\in M$ determined by
$k\cdot u_{0}=u_{0}\cdot\lambda(k)\; \forall k\in I_{x_{0}}$ if
$u_{0}\in\pi^{-1}(x_{0})$.

Wang's theorem \cite{k4872,k0929} on connections that are invariant
under actions transitive on the base manifold has been adapted to spherically
symmetric space-time manifolds by Brodbeck and Straumann \cite{k5281}. They
show that in a Schwarzschild type coordinate system $(t,r,\theta,\phi)$ and
the metric
\leqn{metric}{g = -N S^{2} dt^{2}+
  N^{-1}dr^{2}+r^{2}(d\theta^{2}+\sin^{2}\theta d\phi^{2})} 
a gauge can always be chosen such that the \YM-connection form is locally
given by
\eqn{
connx}{A=\tilde{A}+\hat{A}
} 
where $\tilde{A}$ is a 1-form on the quotient space parametrized by the $r$
and $t$ coordinates and
\leqn{Ahat}{
  \hat{A} = \Lambda_{1} d\theta +
            ( \Lambda_{2}\sin\theta + \Lambda_{3}\cos\theta)d\phi
} 
where $\Lambda_{3}=-\ihalf \Lo$ is the constant isotropy generator and
$\Lambda_{1}$ and $\Lambda_{2}$ are functions of $r$ and $t$ that satisfy
\leqn{wang}{ 
[\Lambda_{2},\Lambda_{3}]=\Lambda_{1} \AND
[\Lambda_{3},\Lambda_{1}]=\Lambda_{2}. 
}
Since we only consider static fields we can assume that $\Lambda_{1}$ and
$\Lambda_{2}$ depend only on $r$.  Moreover, we will also concentrate on the
``magnetic'' case and assume that the part $\tilde{A}$ of the gauge potential
which contributes ``electric'' or ``Coulomb'' terms vanishes, i.e. we put
\eqn{Coulomb}{
\tilde{A}=0.
}
This condition is not as restrictive as it seems. For, as proved in
\cite{k4295}, it also follows in the regular case (defined below) if the
field is smooth at the center $r=0$ and falls off sufficiently fast at
infinity.

So far we still have infinitely many possible actions of $SU(2)$ on the
principal bundle, namely one for each element in the intersection
$\overline{W(S)}\cap I$ of the fundamental Weyl chamber and the integral
lattice $I := \kernel(\exp\restr{\h})$. However, since we want the
\YM-connection to be regular also at a center ($r=0$, defined as a connected
set of fixed points of the $SU(2)$-action on $M$) and/or the \YM-field to fall
off in an asymptotic region (have no magnetic charge according to
\cite{k4295}) we must have
\leqn{center}{
    [\Omega^{0}_{1},     \Omega^{0}_{2}]     =\Lambda_{3} 
    \quad\text{and/or}\quad 
    [\Omega^{\infty}_{1},\Omega^{\infty}_{2}]=\Lambda_{3}
}
where 
\eqn{Omdef}{ 
\Omega^{0,\infty}_{i} := \lim_{r\ra0,\infty}\Lambda_{i}(r),\quad(i=1,2).
}
In other words, in these limits there must exist a Lie algebra homomorphism of
$\su(2)$ into $\g_{0}$. This is shown most easily by observing that the
Einstein equations would otherwise lead to infinite pressure or density at a
center.

Since $\Lambda_{3}$ is constant, however, equations \eqref{center} represent
not only conditions on $\Lambda_{1}(r)$ and $\Lambda_{2}(r)$, but also on
$\Lambda_{3}$ and hence on $\Lo$ which must now be the generating (or
defining) vector of an $\sL(2)$ (i.e. $A_{1}$) subalgebra of $\g$. (If both
limits exist it then also follows that there must be an automorphism of $\g$
taking $\Omega^{0}_{i}$ into $\Omega^{\infty}_{i}$.)  The set of these
so-called \Aonev{s}, however, is finite (and in one-to-one correspondence with
conjugacy classes of $\sL(2)$ subalgebras). It has been studied and tabulated
by Mal'cev \cite{k6436} and Dynkin \cite{k4779} and is described by so-called
weighted Dynkin diagrams (called \emph{characteristics} in \cite{k4779}),
where to each simple root in the diagram is associated a number from the set
$\{0,1,2\}$. (See \cite{k6494} for a more recent exposition).  These numbers
represent the values of the simple roots on the generating vector $\Lo$ chosen
such that it lies in $\overline{W(S)}$.

Thus these tables serve as a classification of all the spherically symmetric
``magnetic'' \EYM\ models which are regular at the center and/or obey the
standard fall-off conditions at infinity for any given compact gauge group.

\sect{eqs}{Field equations and reduction of the regular models}

The field equations are well known. We state them here in a form following
\cite{k5281} for the static regular case only, where $\Lo$ is an
$A_{1}$-vector.  Let the space-time metric $g$ be given by \eqref{metric} and
the \YM-potential $A=\hat{A}$ by \eqref{Ahat}. Define, in addition to $\Lo$,
\eqn{lapm}{
\Lpm := \mp \Lambda_{1} - i \Lambda_{2}
} 
so that the Wang equations \eqref{wang} become 
\leqn{nwang}{ 
[\Lo,\Lpm] = \pm 2 \Lpm.  
}
Then $\Lp(r)$ and $\Lm(r)\}$ are $\g$-valued functions, $\Lo$ a (constant)
vector in the fundamental Weyl chamber of $\h$ and $\{\Lo,\Lp,\Lm\}$ is a
standard triple in the limit $r\ra0$ or $r\ra\infty$ for the Lie algebra $\g$.
Now $\h$ is the Cartan subalgebra of the complexified Lie algebra $\g$, i.e.
$\h=\h_{0}\oplus i\h_{0}$, where $\h_{0}$ is the real Cartan subalgebra of a
compact real form $\g_{0}$ of $\g$, and we choose conventions such that the
conjugation operator $c:\g\ra\g$ satisfies $c(X+iY) = X-iY\; \forall\;
X,Y\in\g_{0}$. Then 
\leqn{lam}{
\Lm=-c(\Lp)
} 
so that the dependent variables consist only of $N$, $S$ and the components of
$\Lp$.

The field equations now reduce to
\lgath{feq}{
m' = (NG + r^{-2}P),                         \label{feq1}\\
S^{-1}S' = 2 r^{-1}G,                        \label{feq2}\\
r^{2}N \Lp'' + 2(m-r^{-1}P)\Lp' + \Fc = 0,   \label{feq3}\\
[\Lp,\Lm'] - [\Lp',\Lm] = 0                  \label{feq4}
}
where ${}' := d/dr$ and
\lgath{vardefs}{
N   =: 1 - \frac{2m}{r},  \quad 
G   := \half (\Lp',\Lm'), \quad P := -\half (\Fh,\Fh), \notag \\
\Fh := \ihalf(\Lo-[\Lp,\Lm]),      \label{vardefs4} \\
\Fc := -i[\Fh,\Lp].                \label{vardefs5}
}
Here $(,)$ is an invariant inner product on $\g$. It is determined up to a
factor on each simple component of a semi-simple $\g$ and induces a norm $|.|$
on (the Euclidean) $\h$ and therefore its dual. We choose these factors so
that $(,)$ is a positive multiple of the Killing form on each simple
component. If they are chosen such that the length of the long simple roots
are all 1 then the equations will agree with those in \cite{hka28} for the
principal $SU(n)$ case.

Note that $G\ge0$ and also $P\ge0$.  This follows from \eqref{lam},
$c(\Fh)=\Fh$, and the fact that $\hip{X}{Y} := -(c(X),Y)$ is a Hermitian inner
product on $\g$ (cf. \eqref{HIP}).  Energy density, radial and tangential
pressure are then given by

\leqn{ener}{
 4\pi e          = r^{-2}(NG+r^{-2}P), \quad
 4\pi p_{r}      = r^{-2}(NG-r^{-2}P), \quad
 4\pi p_{\theta} = r^{-4}P.            
}  
We now choose a Chevalley-Weyl basis of $\g$ using mostly the notation of
\cite{k5157}. Let $R$ be the set of roots in $\h^{*}$,
$S=\{\alpha_{1},\ldots,\alpha_{\ell}\}$ a base of $R$ ($\ell$ being the rank
of $\g$), define
\eqn{humph}{ 
\langle \alpha,\beta\rangle := 
\frac{2(\alpha,\beta)}{|\beta|^{2}}, 
}
\eqn{tal}{
(\mathbf{t}_{\alpha},X):=\alpha(X)\;\forall\; X\in\h,
}
and 
\eqn{h}{ 
\hb_{\alpha}:=\frac{2\mathbf{t}_{\alpha}}{|\alpha|^{2}}. 
}
Then 
$\{ \hb_{i}:=\hb_{\alpha_{i}}, \eb_{\alpha}, \eb_{-\alpha} | 
    i=1,\ldots,\ell, \alpha\in R \}$ 
is a basis of $\g$ corresponding to the decomposition
\eqn{decomp}{ 
\g = \h \oplus \bigoplus_{\alpha\in R^{+}}
                    (\g_{\alpha}\oplus\g_{-\alpha}) 
}
($R^{+}$ being the set of positive roots with respect to the base $S$)
for which we choose the conventions
\leqn{chev}{
[\eb_{\alpha},\eb_{-\alpha}] = \hb_{\alpha},\quad
[\eb_{-\alpha},\eb_{-\beta}] = -[\eb_{\alpha},\eb_{\beta}],\quad
(\eb_{\alpha},\eb_{-\alpha}) = \frac{2}{|\alpha|^{2}}.
}
Now it follows directly \cite{k4779} from the defining relations 
\eqn{sl2eq}{ 
[\eb_{0},\eb_{\pm}] = \pm2\eb_{\pm}, \quad
              [\eb_{+},\eb_{-}]=\eb_{0} 
}
of an $\sL(2)$-subalgebra span\{$\eb_{0},\eb_{\pm}$\} of $\g$, with
the help of
\eqn{hea}{
[\mathbf{h},\eb_{\alpha}]=\alpha(\mathbf{h})\eb_{\alpha},
} 
that
$\eb_{0}$ can only be an \Aonev\  provided
\eqn{A1vec}{ 
\alpha(\eb_{0}) = 2 \quad\text{for some\ } \alpha\in R.
}
Thus, if we let
\leqn{laz}{ 
\Lo = \sum_{i=1}^{\ell}\lambda_{i}\hb_{i} \in \h. 
}
then equations \eqref{nwang} imply that
\eqn{lap}{ 
\Lp(r) = \sum_{\alpha\in S_{\lambda}} w_{\alpha}(r) \eb_{\alpha}
}
where
\leqn{Slam}{
S_{\lambda} := \{\alpha\in R\, | \, \alpha(\Lo)=2 \} 
}
is a set of roots depending only on the homomorphism $\lambda$ or,
equivalently, on the coefficients $\lambda_{i}$ in \eqref{laz}.

Similarly we have
\leqn{clam}{ 
\Lm(r) = \sum_{\alpha\in S_{\lambda}} v_{\alpha}(r)
  \eb_{-\alpha}, 
}
but by \eqref{lam} and the fact that complex conjugation maps 
\eqn{conj}{
c:\hb_{i}\mapsto -\hb_{i}, \eb_{\alpha}\mapsto -\eb_{-\alpha}
} 
it follows that
\eqn{cw}{ 
v_{\alpha}(r) = \bar{w}_{\alpha}(r). 
} 
Our system is thus determined once the two real functions $m(r)$ and $S(r)$
and the complex functions $w_{\alpha}(r)$ for all $\alpha\in S_{\lambda}$ are
known.

If we now substitute \eqref{clam} into equations \eqref{feq1} to
\eqref{vardefs5} we need to calculate the Lie brackets between the various
$\eb_{\alpha}$ for $\alpha\in S_{\lambda}$. In general, this may produce many
more equations than dependent variables. On the other hand the \YM-potential
$\hat{A}$ determined by $\Lp$ still contains some gauge freedom. It is not
known, at present, whether there is any systematic method to solve this
system of equations.

However, as Brodbeck and Straumann \cite{k4295} have observed, there are
special symmetry actions for which this system of equations is much simpler,
in fact, very similar to the principal $SU(n)$ case. This happens when $\Lo$
is a vector in the \emph{open} fundamental Weyl chamber of $\h$. They call
these models generic, but since, as we will see, they are really a small
minority of all possible ones we will call them \emph{regular}.

In the following $\Lambda_{0}$ is not required to be an \Aonev. 
\begin{thm}[Brodbeck/Straumann \cite{k4295}]\mlabel{pisys}
If $\Lo$ is in the open Weyl chamber $W(S)$ then the set
$S_{\lambda}$ is a $\Pi$-system, i.e. satisfies 
\begin{itemize}
\item[(i)] if $\alpha, \beta\in S_{\lambda}$ then $\alpha-\beta \notin R$,
\item[(ii)] $S_{\lambda}$ is linearly independent 
\end{itemize}
and is therefore the base of a root system $R_{\lambda}$ which generates a Lie
subalgebra $\g_{\lambda}$ of $\g$ spanned by
$\{\hb_{\alpha},\eb_{\alpha},\eb_{-\alpha}|\alpha\in R_{\lambda}\}$.\\
Moreover, if $\h_{\lambda}:=\mathrm{span}\{\hb_{\alpha}|\alpha\in
S_{\lambda}\}$ and $\h_{\lambda}^{\perp}:=\bigcap_{\alpha\in
  S_{\lambda}}\kernel\,{\alpha}$ then
\eqn{laOdec}{
  \h=\h_{\lambda}^{\shortparallel}\oplus \h_{\lambda}^{\perp} \AND 
              \Lo = \Lo^{\shortparallel}+\Lo^{\perp} \quad \text{with}
  \quad \Lo^{\shortparallel} = \sum_{\alpha\in R_{\lambda}^{+}}\hb_{\alpha}. 
}
If $\Lambda_{0}$ is an \Aonev\ then $\Lo^{\perp}=0$ (but $\h_{\lambda}^{\perp}$ need not be trivial).
\end{thm}

In particular, $\Lo^{\shortparallel}$ is twice the lowest weight vector of
$\h_{\lambda}$ and we have by the definitions of $S_{\lambda}$ and $\h_{\lambda}^{\perp}$
\eqn{weight}{ 
\alpha(\Lo^{\shortparallel}) = 2 \AND \alpha(\Lo^{\perp})=0
  \quad \forall\;\alpha\in S_{\lambda}. 
}
We will from now on only consider the regular case. 

First, $\Lp$ can be treated as a $\g_{\lambda}$-valued function,
\eqn{lapn}{ 
\Lp(r) = \sum_{j=1}^{\ell_{\lambda}} w_{j}(r) \tilde{\eb}_{j} 
} 
where now $\{\tilde{\alpha}_{1},\ldots,\tilde{\alpha}_{\ell_{\lambda}}\}$ is
the base of $S_{\lambda}$ and $\tilde{\eb}_{j}:=\eb_{\tilde{\alpha}_{j}}$.
Moreover, $\Lo^{\shortparallel} = \sum_{j}^{\ell_{\lambda}}
\lambda^{\shortparallel}_{j} \tilde{\hb}_{j}$ with
$\tilde{\hb}_{j} :=\hb_{\tilde{\alpha}_{j}}$.

Then, by \eqref{vardefs4} noting also that $\tilde{\alpha_{j}}(\Lo^{\perp}) =
0$,
\leqn{Fh}{ 
\Fh = \frac{i}{2} \left( 
    \DS\sum_{j=1}^{\ell_{\lambda}} \lambda^{\shortparallel}_{j}\tilde{\hb}_{j} + \Lo^{\perp} - 
    \left[ \DS\sum_{j=1}^{\ell_{\lambda}}w_{j}\tilde{\eb}_{j},
           \DS\sum_{k=1}^{\ell_{\lambda}}\bar{w}_{k}\tilde{\eb}_{k} \right] \right) 
    = \frac{i}{2}\left( \DS\sum_{j=1}^{\ell_{\lambda}}(\lambda^{\shortparallel}_{j}-|w_{j}|^{2}) \tilde{\hb}_{j} + \Lo^{\perp} \right)
}
where \eqref{chev} was used and the fact that differences of two simple roots are
not roots which implies that 
\leqn{eab}{   
  [\eb_{\alpha},\eb_{-\beta}] = 0 \quad \forall\;\alpha,\beta\in S_{\lambda},\;
  \alpha\neq\beta.  
}
Substituting this expression into \eqref{vardefs5} gives
\leqn{Fc}{ 
  \Fc = \half \sum_{j,k=1}^{\ell_{\lambda}}w_{j}\langle
  \tilde{\alpha}_{j},\tilde{\alpha}_{k}\rangle
  (\lambda^{\shortparallel}_{k}-|w_{k}|^{2}) \tilde{\eb}_{j} + \half
  \sum_{j=1}^{\ell_{\lambda}} w_{j}[\Lo^{\perp},\tilde{\eb}_{j}]. 
}

But since $[\Lo^{\perp},\tilde{\eb}_{k}] = \tilde{\alpha}_{k}(\Lo^{\perp})\tilde{\eb}_{k} = 0 $, in
view of the definition of $S_{\lambda}^{\perp}$, the last term vanishes.
Equation \eqref{feq3} therefore becomes
\eqn{nfeq3}{ 
r^{2} N w_{j}'' + 2(m-r^{-1}P)w_{j}' + \half
  \DS\sum_{k=1}^{\ell_{\lambda}} w_{j} c_{jk} (\lambda^{\shortparallel}_{k}-|w_{k}|^{2}) = 0 
}
where we have introduced
\eqn{cmat}{ 
c_{jk} := \langle \tilde{\alpha}_{j}, \tilde{\alpha}_{k} \rangle
} 
for the Cartan matrix of $\g_{\lambda}$ and where now
\lgath{PG}{ 
  P = {\textstyle\frac{1}{8}}
  \sum_{j,k=1}^{\ell_{\lambda}}(\lambda^{\shortparallel}_{j}-|w_{j}|^{2})h_{jk}(\lambda^{\shortparallel}_{k}-|w_{k}|^{2}) +
  |\Lo^{\perp}|^{2} \quad \text{with\ } 
   h_{jk}:= \frac{2 \langle \tilde{\alpha}_{j}, \tilde{\alpha}_{k}\rangle }
                  { |\tilde{\alpha}_{j}|^{2} },
  \label{PG1}\\
  G = \sum_{j=1}^{\ell_{\lambda}}\frac{|w_{j}'|^{2}}
       { |\tilde{\alpha}_{j}|^{2} }. \label{PG2}
}
Finally, \eqref{feq4} simply becomes in view of \eqref{chev}
\leqn{nfeq4}{
  \sum_{j,k=1}^{\ell_{\lambda}}\left(w_{j}\bar{w}_{k}'-w_{j}'\bar{w}_{k}\right)[\eb_{\tilde{\alpha}_{j}},\eb_{-\tilde{\alpha}_{k}}]
  = \sum_{j=1}^{\ell_{\lambda}} (w_{j}\bar{w_{j}}' - w_{j}'\bar{w}{j}) \tilde{\hb}_{j}
  = 0 
}
so that the phase of $w_{j}$ is constant and may be chosen to be zero by a
gauge transformation. One can thus assume that the $w_{j}(r)$ are
real-valued functions.

It remains to determine the subalgebra $\g_{\lambda}$ for a given \Aonev\ 
$\Lo$ in the open fundamental Weyl chamber.

First, we note that for a semisimple group for which the Cartan subalgebra $\h$ splits into an orthogonal sum $\h = \bigoplus \h_{k}$  the decomposition in Theorem \ref{pisys} splits into corresponding decompositions of each of the $\h_{k}$. So we need only investigate the regular actions of simple Lie groups.

Now the \Aonev\ in the Cartan subalgebra $\h$ of an semisimple Lie algebra
$\g$ is uniquely given by the numbers 
\leqn{char}{ 
\mathbf{\chi} =
  (\chi_{1},\ldots,\chi_{\ell}) :=
  \bigl(\alpha_{1}(\Lo),\ldots,\alpha_{\ell}(\Lo)\bigr), 
} 
called the \emph{characteristic} in \cite{k4779}. It is known
\cite{k4779,k6494} that if $\Lo$ is in the closed fundamental Weyl chamber
then $\chi_{k}\in \{0,1,2\}$, and all possible such characteristic have been
found and tabulated. It is clear from the definition of $S_{\lambda}$ in
\eqref{Slam} that $\chi_{k}=2 \;\forall\;k$ for $\h_{\lambda}$. Such \Aonev{s}
define \emph{principal} $A_{1}$-subalgebras and thus \emph{principal actions}
of $SU(2)$ on the bundle. We now have
\begin{thm}\mlabel{classreg}{\ }\vspace{-5pt}  
\begin{itemize}
\item[(i)] The possible regular $A_{1}$-subalgebras of simple Lie algebras
  consist of the principal subalgebras of all Lie algebras $A_{\ell}$,
  $B_{\ell}$, $C_{\ell}$, $D_{\ell}$, $E_{\ell}$, $F_{4}$ and $G_{2}$ and of
  those subalgebras of $A_{\ell}=\sL(\ell+1)$ with even $\ell$ corresponding
  to partitions $[\ell+1-k,k]$ for any integer $k=1,\ldots,\ell/2$ or,
  equivalently, characteristic $(22..211..112..22)$ ($2k$ `$1$'s in the
  middle, `$2$'s in all other positions).
\item[(ii)] The Lie algebra $\g_{\lambda}$ is equal to $\g$ in the principal
  case, and for $A_{\ell}$ with even $\ell$ equal to $A_{\ell-1}$ for $k=1$
  and to $A_{\ell-k}\oplus A_{k-1}$ for $k=2,\ldots,\ell/2$.
\item[(iii)] In the principal case $\h_{\lambda}^{\shortparallel}=\h$. For all
  $A_{1}$-subalgebras of $A_{\ell}$ with even $\ell$ the orthogonal space
  $\h_{\lambda}^{\perp}$ is one-dimensional.
\end{itemize}
\end{thm}
\begin{proof}
  Part (i) follows quite easily from the discussion and the tables in
  \cite{k6494} (Sections 5.3 and 4.4).\\ 
  For part (ii) that $\h_{\lambda}=\h$ in the principal case is
  obvious. To compute $S_{\lambda}$ for a given $\ell=2m$ and given $k>0$ note
  that all positive roots of $A_{\ell}$ are of the form
  $\sum_{p=j}^{k}\alpha_{p}$ for $1\leq j\leq k\leq 2m$ so that using that
  $\alpha_{i}(\Lo)=2$ for $i=1,\ldots,m-k$ and $i=m+k+1,\ldots,2m$ and
  $\alpha_{i}(\Lo)=1$ otherwise one sees that
  \leqn{decSl}{ 
    S_{\lambda}=\bigcup_{i=1}^{m-k}\alpha_{i} \cup
     \bigcup_{j=1}^{2k-1}(\alpha_{m-k+j}+\alpha_{m-k+j+1})\;\; \cup
     \bigcup_{i=m+k+1}^{2m}\alpha_{i}.
  }
  Recalling that for $A_{\ell}$
  $$\langle \alpha_{i},\alpha_{j}\rangle = \left\{
      \begin{array}{cl}
        2  & \text{if\ }i=j   \\ 
        -1 & \text{if\ }|i-j|=1\\ 
        0  & \text{otherwise}
       \end{array}
     \right.$$
     it is seen immediately that the Cartan matrix for $S_{\lambda}$
     is the one for $A_{\ell-1}$ if $k=1$ while it takes a simple reordering
     of the roots to verify the statement in (ii) for $k=2,\ldots,\ell/2$.\\
     (iii) That $\h_{\lambda}^{\shortparallel}=\h$ in the principal case is
     obvious from the definition and that $\dim S_{\lambda}^{\perp}=1$ follows
     from the observation that $\alpha(X)=0\;\forall\; \alpha\in S_{\lambda}$
     amounts to $2m-1$ linearly independent equations according to
     \eqref{decSl}.
\end{proof}
In summary, we have shown that all regular models can be reduced to those
with the principal action for semisimple gauge groups. Also the term
$\Lo^{\perp}$ occuring in \eqref{Fh}, \eqref{Fc} and \eqref{PG1} can now be
dropped.

\sect{sol}{Constructing local solutions regular at the center, horizon or at
infinity}

So far we have shown that the static spherically symmetric and magnetic EYM
equations for the regular action reduce to those for the principal action
for semi-simple gauge groups. (We now drop the index $\lambda$ from $\g$,
$\h$, etc.) They consist of \eqref{feq2}, which can be integrated easily once
the other equations are solved, and
\lgath{neqs}{ m' = (NG + r^{-2}P), \label{neqs1}\\
  r^{2} N w_{j}'' + 2(m-r^{-1}P)w_{j}' + \half
  \DS\sum_{k=1}^{\ell} w_{j} c_{jk}(\lambda_{k}-w_{k}^{2}) = 0\label{neqs2} 
}
where the $w_{k}$ are real-valued functions of $r$, $(c_{ij}):=(\langle
\alpha_{j}, \alpha_{k} \rangle)$ is the Cartan matrix of the reduced structure
group, and
\lalign{nPG}{ P &= {\textstyle\frac{1}{8}}
  \sum_{j,k=1}^{\ell} (\lambda_{j}-w_{j}^{2}) h_{jk} (\lambda_{k}-w_{k}^{2}) 
  \label{nPG1}\\
  G &= \sum_{j=1}^{\ell}\frac{{w_{j}'}^{2}} { |\alpha_{j}|^{2} } \label{nPG2}\\
  h_{jk} &= \frac{2 c_{jk}}{|\alpha_{j}|^{2}}\label{nPG3}\\
  \lambda_{j} &= 2 \sum_{k=1}^{\ell}(c^{-1})_{jk}\label{nPG4}
}
The expressions for components $\lambda_{k}$ of the \Aonev\ $\Lo$ follow from
\eqref{char} and the fact that for the principal action
$\chi_{k}=2\;\forall\;k$.

In this section we will discuss the general problem of finding solutions that are regular at the center or at the horizon and have an appropriate fall off as $r\ra\infty$. Proofs will be given later in sections \ref{BFM1} and \ref{BFM2}.

Equations \eqref{neqs1} and \eqref{neqs2} are very similar to the
corresponding ones in the principal $SU(n)$ case analyzed in detail in
\cite{hka28}. So we can expect most of those results to generalize. First of
all, when the dependent variables $m$ and $w_{k}$ are expanded in power series
in terms of $r$ at $r=0$, in terms of $r-\rh$ at $r=\rh$, and in terms of
$r^{-1}$ at infinity (under the assumption that all the quantities are finite
in these limits) then \eqref{neqs1} and \eqref{neqs2} yield a system of
algebraic equations. For example, at $r=0$ with $f(r)=\sum_{k=0}^{\infty}
f_{k}r^{k}$ we find
\lgath{expans}{
  m_{k+1} = \frac{1}{k+1}\left( G_{k}+P_{k+2}-2\sum_{h=2}^{k-2}m_{k-h}G_{h}
   \right), \label{expans1}\\
  \sum_{j=1}^{\ell}\bigl( A_{ij} - k(k+1)\delta_{ij} \bigr) w_{j,k+1} =
  b_{i,k} \label{expans2} 
} 
for $k=0,1,2,\ldots$ where
\leqn{Aij}{ A_{ij} :=  w_{i,0}c_{ij}w_{j,0} }
and the $b_{i,k}$ are complicated expressions involving lower order terms.
For the lowest order terms we find
\leqn{lowest}{ m_{0}=m_{1}=m_{2}=0,\quad w_{i,0}^{2}=\lambda_{i},\quad w_{i,1}=0. }
That $r=0$ is a singular point for the system \eqref{neqs1},\eqref{neqs2}
manifests itself in the fact that the initial data at $r=0$ for regular
solutions are not simply the values of the functions $m$, $w_{i}$ and $w'_{i}$
but that some of these values are restricted like in \eqref{lowest} and some
higher order coefficients in the power series for the $w_{i}$ remain
arbitrary, namely for those orders $k$ for which the matrix
$\mathbf{A}=(A_{ij})$ has eigenvalue $k(k+1)$. It now turns out that the
eigenvalues of $\mathbf{A}$ are precisely of this form for certain integer
values of $k$. In fact, for the simple Lie algebras we can calculate the
spectrum directly from the Cartan matrix and find the values given in
Table~\ref{Aspec}. The proof for the classical Lie algebras of arbitrary rank follows from the properties of the root system and results at the end of section \ref{BFM1}.

\begin{center}
\begin{table}[ht]
\caption{The eigenvalues of the coefficient matrix $\mathbf{A}$ 
for the simple Lie algebras are given by the set $\text{spec}(\mathbf{A}) 
= \{k(k+1) | k\in\mathcal{E} \}$. For the classical Lie algebras the table 
entry gives $k_{j}$ for $j=1,2,\ldots=\ell=\text{rank}(\g)$.  Note that 
$k=1$ belongs to $\Ec$ for all Lie algebras.} \label{Aspec}
$$
\begin{array}{||c|c||}
\hline\hline
\text{Lie algebra} & \Ec \\
\hline\hline
A_\ell             & j \\
\hline
B_\ell             & 2j-1 \\
\hline
C_\ell             & 2j-1 \\
\hline
D_\ell             &  \left\{\begin{matrix} 
                      2j-1     & \text{\ if\ } j\leq (\ell+2)/2 \\ 
                      \ell-1   & \text{\ if\ } j=(\ell+2)/2      \\
                      2j-3     & \text{\ if\ } j>(\ell+2)/2
                      \end{matrix}\right. \\
\hline
E_6                & 1,4,5,7,8,11 \\
E_7                & 1,5,7,9,11,13,17 \\
E_8                & 1,7,11,13,17,19,23,29 \\
\hline
F_4                & 1,5,7,11\\
\hline
G_2                & 1,5\\
\hline\hline
\end{array}
$$
\end{table}
\end{center}

The eigenspaces for the simple Lie algebras are all onedimensional except for
$D_{\ell}$ where certain 'middle' eigenvalues occur twice. For semisimple Lie
algebras the matrix $\mathbf{A}$ will be a direct sum of those for the simple
components and thus may have multiple eigenvalues.

It is now clear that a formal power series solution of equations \eqref{neqs1}
and \eqref{neqs2} is well defined and contains $\ell$ free parameters provided
equation \eqref{expans2} can be solved, i.e. provided the vector
$\mathbf{b}_{k}:=(b_{1,k},\ldots,b_{\ell,k})$ lies in the left kernel of
$\left(\mathbf{A}-k(k+1)\id\right)$. Since $\mathbf{b}_{k}$ is a very
complicated expression this is cumbersome to prove in general. In \cite{hka28}
the proof for $G=SU(n)$ was achieved using properties of a class of orthogonal
polynomials, an approach that does not easily generalize to other groups. In
sections \ref{BFM1} and \ref{BFM2} we present a proof that depends directly on
the root structure of the Lie algebra $\g$ treated as an $\Sl{2}$-module.

The structure of the recursion relations for the power series of regular
solutions in $r^{-1}$ at infinity is very similar to the one at $r=0$. At the
remaining singular point of \eqref{neqs2}, namely at a regular horizon
($N(\rh)=0, N'(\rh)>0$), however, the only conditions on initial
values turn out to be some inequalities.

Calculating the formal power series is indeed necessary to start off numerical
integration when searching for global regular solutions. For an existence and
uniqueness proof, however, it is more convenient to recast the equations in a
form to which the following (slight generalization of a) theorem by Breitenlohner, Forg\'acs and Maison \cite{k5428}
applies.
\begin{thm} \mlabel{BFM}
The system of differential equations
\lalign{bfmeq}{
t \frac{du_{i}}{dt} &= t^{\mu_{i}}f_{i}(t,u,v) & i=1,\ldots,m \\
t \frac{dv_{j}}{dt} &= -h_{j}(u)v_{j} + t^{\nu_{j}}g_{j}(t,u,v) & j=1,\ldots,n
}
where $\mu_{i}$, $\nu_{j}$ are integers greater than 1, $f_{i}$ and $g_{j}$ 
analytic functions in a neighborhood of $(0,c_{0},0)\in \RE^{1+m+n}$, and 
$h_{j}:\RE^{m}\ra\RE$ functions, positive in a neighborhood of 
$c_{0}\in\RE^{m}$, has a unique analytic solution 
$t\mapsto(u_{i}(t),v_{j}(t))$ such that
\leqn{bfmsol}{ 
u_{i}(t) = c_{i} + O(t^{\mu_{i}}) \AND  v_{j}(t) = O(t^{\nu_{j}}) 
}
for $|t|<R$ for some $R>0$ if $|c-c_{0}|$ is small enough. Moreover, the solution depends analytically on the parameters $c_{i}$.
\end{thm}
\begin{proof}
  By the standard method solving the differential equation with initial data
  is replaced by finding a fixed point for the map $T:
  (u,v)\mapsto(\tilde{u},\tilde{v})$ with
\lalign{inteq}{
\tilde{u}_{i}(t) &= c_{i} + \int_{0}^{t}\tau^{\mu_{i}-1}f_{i}[\tau,u(\tau),v(\tau)]\,d\tau \\
\tilde{v}_{j}(t) &= t^{-\kappa_{j}}\int_{0}^{t}\tau^{\kappa_{j}\nu_{j}-1}\hat{g}_{i}[\tau,u(\tau),v(\tau)]\,d\tau
}
where $\kappa_{j}:=h_{j}(c)$ and $\hat{g}(j)(t,u,v) := g_{j}(t,u,v) -
t^{-\nu_{j}}[h_{j}(u)-h_{j}(c)]v_{j}$. To show that $T$ is a contracting map
on a suitable Banach space one can use a method very similar to the one in
\cite{hka28}.
\end{proof}
To bring the system \eqref{neqs1} and \eqref{neqs2} into a form that satisfies
the hypotheses of theorem \ref{BFM} it is necessary to make a suitable
transformation of the variables $m$ and $w_{j}$. The proofs that this can be done
are basically equivalent to showing that the formal power series
exist and are given in section \ref{BFM2}. We then have
\begin{thm} \mlabel{exist0}
  The system \eqref{neqs1} and \eqref{neqs2} has an analytic solution for
  small $r$ of the form
\leqn{wu0}{ 
   w_{i}(r) = w_{i,0} + \sum_{j=1}^{\ell} C_{ij} r^{k_{j}+1}u_{j}(r),
   \quad i=1,\ldots,\ell 
}
where $C=(C_{ij})$ is a nonsingular matrix whose $j$-th column is an
eigenvector to eigenvalue $k_{j}(k_{j}+1)$ of the matrix $\mathbf{A}$. The
solution is uniquely determined by the initial values $u_{j}(0) = \beta_{j}$
for arbitrary $\beta_{j}$. The function $m(r)$ is then determined and
satisfies $m(r) = O(r^{3})$ for small $r$.
\end{thm}
Note that the $w_{i,0}$ are determined up to the sign by \eqref{lowest}.
From \eqref{origbfm21a}, we see that  solutions from theorem \ref{exist0} satisf
y $P = \text{O}(r^4)$
and $G = \text{O}(r^2)$. It follows that for these solutions all physical
quantities such as the pressure and mass density are finite at $r=0$.

The situation is rather similar for solutions analytic in $r^{-1}$ near infinity. We have, with the same matrix $C$,
\begin{thm} \mlabel{existInf}
  The system \eqref{neqs1} and \eqref{neqs2} has an analytic solution for small
  $z=r^{-1}$ of the form
\lalign{wuInf}{ 
   w_{i}(r) &= w_{i,\infty} + \sum_{j=1}^{\ell} C_{ij} r^{-k_{j}}u_{j}(r^{-1}),
   \quad i=1,\ldots,\ell, \label{wuInf1} \\
   m(r)     &= m_{\infty} + O(r^{-1}) \label{wuInf2}
}
The solution is uniquely determined by
the initial values $u_{j}(0) = \alpha_{j}$ and $m_{\infty}$ for arbitrary $\alpha_{j}$ and $m_{\infty}$. 
\end{thm}
Again $w_{i,\infty}$ is determined up to the sign by
$w_{i,\infty}^{2}=\lambda_{i}$. An overall sign in $w_{i}(r)$ does not affect
the Yang-Mills field nor the geometry and physics. But $w_{i,0}$ and
$w_{i,\infty}$ may have the same or different signs for global solutions.

Finally we have the corresponding theorem for local solutions near a regular
horizon.
\begin{thm} \mlabel{existHor}
  The system \eqref{neqs1} and \eqref{neqs2} has a solution analytic in
  $t=r-\rh$ for small $t$ at a regular horizon, i.e. where
  $N(\rh)=0$ and $N'(rH)>0$. The solution is uniquely
  determined by the values of $w_{j}(\rh)$ which must be chosen such that
\lgath{wuHor}{ 
   N'(\rh) = \frac{1}{\rh}-\frac{2}{\rh^{3}}P(\rh) > 0\\
\intertext{or, equivalently,}
   2 P(\rh) = \quart \sum_{i,j=1}^{\ell}\bigl(\lambda_{i}-w_{i}\bigl(\rh)\bigr)h_{ij}\bigl(\lambda_{j}-w_{j}(\rh)\bigr) < \rh^{2}.   
}
\end{thm}

\sect{elem}{Elementary properties and scaled solutions}

The observation already made in \cite{k4752} for $SU(2)$ and generalized to $SU(n)$ that global solutions, if they exist, must be bounded by their values at infinity (or zero) is easily extended to the regular case for arbitrary $G$.
\begin{thm} \mlabel{bounded}
If a solution $(m,w_{1},\ldots,w_{\ell})$ is defined and $C^{2}$ in the connected outer domain $D:=\{r|0 \leq \rh \leq r < \infty\}$ (where $N(r)>0$) and if
\eqn{asym}{ 
m(r)=m_{\infty}+O(1/r)\AND w_{j}(r)=w_{j,\infty}+O(1/r)\;\text{as}\;r\ra\infty
}
then
\eqn{bnd}{ 
  w_{j}(r)^{2} \leq w_{j,\infty}^{2} = 
  \lambda_{j} \quad\forall\;r\in\overset{\circ}{D}.
}
Moreover, if $G$ is a simple group and $w_{j}(r_{1})=w_{j,\infty}\;\;\text{for
  some\ }j \text{\ and for some\ } r_{1}\in \overset{\circ}{D}$ then
$w_{j}(r)=w_{j,\infty}\;\forall\;r\in D$, $m=\const$, the Yang-Mills field
vanishes, and the metric is the Schwarzschild one.

If $G$ is semisimple and  $w_{j}(r_{1})=w_{j,\infty}\;\;\text{for
  some\ }j \text{\ and for some\ } r_{1}\in \overset{\circ}{D}$ then the field equations reduce to those of the subgroup of $G$ obtained by deleting from the Cartan subalgebra $\h$ of $\g$ the simple component in which $\hb_{j}$ lies.
\end{thm}
\begin{proof}
Let $v_{j}:=w_{j}^{2}$. Then $v_{j}(r)\geq0\;\forall\;r$ and \eqref{neqs2} gives
\leqn{veq}{
2r^{2}N v_{j}v_{j}'' - r^{2}N {v_{j}'}^{2} + 4(m-r^{-1}P)v_{j}v_{j}' + 2v_{j}^{2}\sum_{k=1}^{\ell}c_{jk}(\lambda_{k}-v_{k}) = 0.
}
Let $V_{j}:=\sup_{r\in D} v_{j}(r)$. Then $V_{j}>0$ because the asymptotic
value of $v_{j}(r)$ is $\lambda_{j}>0$. Now assume that $v_{j}(r_{j})=V_{j}$
for some $r_{j}\in \overset{\circ}{D}$. Then $v_{j}(r_{j})$ is an absolute
maximum so that $v'_{j}(r_{j})=0$ and $v''_{j}(r_{j})\leq0$. It follows from
\eqref{veq} that
$\sum_{j=1}^{\ell}c_{ij}\bigl(\lambda_{j}-v_{j}(r_{j})\bigr)\geq0$ which in
view of \eqref{nPG4} is equivalent to
\gath{veq2}{
 \sum_{j=1}^{\ell}c_{ij}v_{j}(r_{j}) \leq 2 \quad\forall\;i \\
\intertext{or}
 v_{i}(r_{i}) \leq 1 + \half\sum_{j\neq i}(-c_{ij})v_{j}(r_{i}) \leq
    1 + \half\sum_{j\neq i}(-c_{ij})\sup_{r\in D}v_{j}(r) \\
\intertext{whence $V_{i} \leq 1 + \half\sum_{j\neq i}(-c_{ij})V_{j}$ or}
 \sum_{j=1}^{\ell}c_{ij}V_{j} \leq 2 \quad\forall\;i.
}
(Note that for all Cartan matrices $c_{ii}=2$ and $c_{ij}\leq0$ if 
$i\neq j$.)

This last set of inequalities, however, can be multiplied with the inverse
Cartan matrix since the latter has only positive entries. Using \eqref{nPG4}
again then gives $0 \leq V_{j} \leq \lambda_{j}\quad\forall\;j$, thus $V_{j}=\lambda_{j}$ since $\lambda_{j}$ is the asymptotic value.

Suppose now that $v_{i}(r_{*})=\lambda_{i}$ for some $i$ and some $r_{*}\in
\overset{\circ}{D}$.  Then we find from \eqref{veq}
\eqn{veq3}{
 r_{*}N(r_{*})v''_{i}(r_{*})=\lambda_{i} \sum_{j\neq i}(-c_{ij})\bigl(\lambda_{j}-v_{j}(r_{*})\bigr) \geq 0
}
which contradicts that $v_{i}(r)$ has a maximum at $r_{*}$ unless
$v''_{j}(r_{*})=0$ and, in the case of a simple Lie algebra, the neighboring
$v_{j}$ also assume their maximal values. (For a simple Lie algebra there is a
$c_{ij}<0$ for some $j\neq i$ for any $i$.) It follows that all
$v_{j}(r_{*})=\lambda_{j}$ for all $j$ for which the root $\alpha_{j}$ is in
the same simple component of $\h^{*}$. However, if all
$v_{j}(r_{*})=\lambda_{j}$ and thus $v_{j}'(r_{*})=0 \;\forall\;j$ then the
initial conditions for the differential equations \eqref{veq} are all trivial
and since $r_{*}$ is not a singular point it follows by the uniqueness of the
solution that it must by the one for which $v_{j}(r)\equiv\lambda_{j}$. It
then also follows from \eqref{neqs1} that $m=\const$ so that the Yang-Mills
field vanishes and the geometry is the one of the Schwarzschild solution.
\end{proof}

Theorem \ref{bounded} shows among other things that for a given semisimple
gauge group $G$ and a given group action (characterized by $\Lo$) there may be
special solutions that reduce the YM-connection to a subgroup of $G$ that is a
product of some of the simple factors of $G$. Somewhat similarly, since the
group $SU(2)$ can be isomorphically imbedded in every compact (simply
connected) semisimple or simple Lie group the Bartnik-Mckinnon solution
\cite{k4752} can be obtained as a special solution for all the models
considered here. The following special BM-solution for arbitrary compact $G$
was already obtained in \cite{k5945}.

Consider the gauge group $G$ and the symmetry group action (characterized by
$\Lo$) fixed and such that $\Lo$ is regular so that the field equations are
given by \eqref{feq2}, \eqref{neqs1} and \eqref{neqs2}. Select any $\Op$ such
that the set $\{\Lo,\Op,\Om\}$ is a standard triple with
$c(\Omega_{+})=-\Omega_{-}$ and let $\Lp(r)=u(r)\Op$ or, equivalently,
$w_{i}(r)= w_{i,\infty}u(r)$. Then the field equations become
\lgath{seqs}{
  m' = g_{0}\bigl( N{u'}^{2} + \half r^{-2}(1-u^{2})^{2} \bigr), \label{seqs1}\\
  r^{2}N u'' + \bigl(2m-g_{0}r^{-1}(1-u^{2})^{2}\bigr)u'+ g_{0} u(1-u^{2}) = 0, \label{seqs2}\\
  S^{-1}S' = 2g_{0}r^{-1}{u'}^{2}. \label{seqs3} } where $g_{0} =
\quart\sum_{i,j}\lambda_{i}h_{ij}\lambda_{j}$.  By introducing a new radial
variable $x:=r{g_{0}}^{-1/2}$ one sees easily that \eqref{seqs1}-\eqref{seqs3}
reduce to the well studied equations for the $SU(2)$-\EYM\ fields.

Since $\Lo$ fixes the conjugacy class of the symmetry group action on the
bundle different choices of $\Op$ will lead to isomorphic gauge connections,
namely reductions of the $G$-connection to an $SU(2)$-connection on the
principal bundle for the particular space-time. They are thus physically
equivalent.

In view of the existence theorems for the $G=SU(2)$ case
\cite{k5061,k5201,k5428} it now follows that the system \eqref{neqs1} and
\eqref{neqs2} always admits some global solutions

\begin{thm}
  There exists a countably infinite family of globally regular solutions of
  the \EYM-equations for any simply connected compact semisimple gauge group
  $G$ on a static spherically symmetric asymptotically flat space-time
  diffeomorphic to $\RE^{4}$. Similarly, for any $\rh>0$ there exists an
  infinite family of asymptotically flat black hole solutions with black hole
  radius $\rh$.
\end{thm}

\sect{BFM1}{The Lie algebra $\g$ as an $\Sl{2}$ submodule}

In this section we collect all of the algebraic results needed to prove
theorems \ref{exist0} and \ref{existInf}. 
Introduce a non-degenerate Hermitian inner product 
$\hip{\;\,}{\;}: \g\, \times \,\g \rightarrow \mathbb{C}$ by
\leqn{HIP}{
\hip{X}{Y} := -(c(X),Y) \quad \forall \;  X , Y  \in \g \; ,
}
recalling that $c:\g \rightarrow \g$ is the conjugation operator
determined by the compact real form $\g_{0}$. 
Then $\hip{\;\,}{\;}$ restricts to a real positive definite
inner product on $\g_{0}$. From the invariance properties of $(\;,\;)$ it
follows that  $\hip{\;\,}{\;}$ satisfies
\alin{hiprel}{
\hip{X}{Y}  &= \overline{\hip{Y}{X}} \; , \\
\hip{c(X)}{c(Y)}  &=  \overline{\hip{X}{Y}} \; , \\
\hip{[X,c(Y)]}{Z}   &= \hip{X}{[Y,Z]} \; 
}
for all $X,Y,Z \in \g$. Treating $\g$ as a $\mathbb{R}$-linear space
by restricting scalar multiplication to multiplication by reals,
we can introduce a positive definite inner product 
$\rip{\;\,}{\;}: \g\, \times \,\g \rightarrow \mathbb{R}$
on $\g$ defined by
\eqn{ripdef}{
\rip{X}{Y} := \text{Re}\hip{X}{Y} \quad \forall \; X , Y \in \g \; .
}
Let $\norm{\;}$ denote the norm induced on $\g$ by $\rip{\;}{\;}$, i.e. 
\leqn{norm}{
\norm{X} = \sqrt{\rip{X}{X}} \quad \forall \; X\in \g \; .
}
From the above properties satisfied by $\hip{\;\,}{\;}$, it 
straighforward to verify that $\rip{\;\,}{\;}$ satisfies
\lalign{rip}{
\rip{X}{Y}  &= \rip{Y}{X} \notag \; , \\
\rip{c(X)}{c(Y)}  &=  \rip{X}{Y} \label{rip} \; , \\
\rip{[X,c(Y)]}{Z}   &= \rip{X}{[Y,Z]} \notag \; 
}
for all $X,Y,Z \in \g$.

Let $\Op,\Om \in \g$ be two vectors such that
\eqn{commutation}{
[\Lo,\Omega_{\pm}] = \pm 2 \Omega_{\pm}\:, \;\; [\Op,\Om] = \Lo \AND c(\Op) = -\Om  \; . 
}
Then $\cspan{\{\Lo,\Op,\Om\}} \cong \Sl{2}$. The dot notation will often
be used to denote the adjoint action of 
 $\cspan{\{\Lo,\Op,\Om\}}$ on $\g$, i.e.
\eqn{ad}{
X.Y := \ad(X)(Y) \qquad \forall\; \text{$X\in \cspan{\{\Lo,\Op,\Om\}}$, $ Y\in\g$.}
}
Because $\Lo$ is a
semisimple element, $\ad(\Lo)$ is diagonalizable and it follows from
$\mathfrak{sl}(2)$-representation theory \cite{k5157} that the eigenvalues are integers.
Let $V_{n}$ denote the eigenspaces of $\ad(\Lo)$, i.e.
\eqn{Vn}{
V_{n} := \{ X\in \g \; | \; \Lo . X = n X \}  \quad n\in \mathbb{Z} \; .
}
It also follows from $\Sl{2}$-representation theory that if
$X\in \g$ is a highest weight vector of the adjoint representation
of $\cspan{\{\Lo,\Op,\Om\}}$ with weight $n$, and we
define  $X_{-1} = 0$, $X_{0} = X$ and
$X_{j} = (1/j!)\Om^{j} . X_{0} \quad (j\geq 0)$, then
\lalign{ladder}{
\Lo . X_{j} & = (n-2j) X_{j} \notag \; ,\\
\Om . X_{j} & = (j+1) X_{j+1} \label{ladder} \; ,\\
\Op . X_{j} & = (n-j+1) X_{j-1} \quad (j\geq 0) \notag \; .
}

\begin{prop} \label{hwv} \mnote{[hwv]}
There exists $M$ highest weight vectors
${\xi^{1},\xi^{2},\ldots,\xi^{M} }$ for the
adjoint representation of {\rm $\cspan{\{\Lo,\Op,\Om\}}$} on $\g$
that satisfy 
\begin{itemize}
\item[(i)] the  $\xi^{j}$ have weights $2k_{j}$ where $j=1,2,\ldots,M$ and
$1 = k_{1} \leq k_{2} \leq \cdots \leq k_{M}$,
\item[(ii)] if $V(\xi^{j})$ denotes  the irreducible submodule of $\g$ generated by
$\xi^{j}$, then the  sum $\sum_{j=1}^{M} V(\xi^{j})$ is direct,
\item[(iii)] if $\xi^{j}_{l} = (1/l!)\Om^{l} . \xi^{j}$ then
\leqn{xiconj}{
c(\xi^{j}_{l}) = (-1)^{l} \xi^{j}_{2k_{j}-l} \; ,
}
\item[(iv)] $M = |S_{\lambda}|$ and the set $\{\xi^{j}_{k_{j}-1} \; | \; j= 1,2,\ldots M\}$ forms
a basis for $V_{2}$ over $\mathbb{C}$.
\end{itemize}
\end{prop}

\begin{proof}
\emph{(i)} and \emph{(ii)}: The conjugation operator $c$ satisfies
\leqn{cbrk}{
c([X,Y]) = [c(X),c(Y)]  \quad \forall \; X,Y \in \g \; .
}
Because  $\Lo \in i\h_{0}$,
\leqn{cLo}{
c(\Lo) = -\Lo.
}
Using \eqref{lam}, \eqref{cbrk}, and \eqref{cLo}, it is easy
to see that
\leqn{cmute}{ 
c \circ \ad(\Omega_{\pm})^{n}  = (-1)^{n}\ad(\Omega_{\pm})^{n}\circ c \;
\text{ \ for every\ } n\in \mathbb{Z}_{\geq 0} \AND 
c \circ \ad(\Lo)  = -\ad(\Lo) \circ c \;. 
}
As usual, define the Casimir operator $\mathcal{C}$ by
\eqn{casim}{
\mathcal{C} = \half \ad(\Lo)^{2} + \ad(\Op)\ad(\Om) + \ad(\Om)\ad(\Op) \; .
}
Then $\g$ can be decomposed as follows \cite{k5772}
\leqn{gdecomp}{
\g = \bigoplus_{p} V(s_{p},v^{p}) \; ,
}
where $V(s_{p},v^{p})$ is a highest weight module generated by the  highest weight vector $v^{p}$ of weight 
$s_{p}$, and it has the property 
\leqn{casimir1}{
\mathcal{C}\restr{V(s_{p},v^{p})} = \left(\half s_{p}^{2}+s_{p}\right)
 \text{id}_{V(s_{p},v^{p})} \quad \forall \; p \;.
}
From \eqref{cmute} it follows that
$\mathcal{C}\circ c = c \circ \mathcal{C}$. Using this result
and \eqref{casimir1}, we see that
\leqn{cinvar}{
c(V(s_{p},v^{p})) \subset V(s_{p},v^{p}) \quad \forall \; p \;.
}
Let $\{s_{p_{1}},s_{p_{2}},\ldots,s_{p_{M}}\}$ be the set of weights from the
decomposition \eqref{gdecomp} that are even and greater than zero. We will
assume that they are ordered so that $s_{p_{1}} \leq s_{p_{2}} \leq \ldots
\leq s_{p_{M}}$. Define $k_{j} = s_{p_{j}}/2$. Then the
$k_{j}$ are positive integers that satisfy $k_{1} \leq,k_{2} \leq \ldots \leq
k_{M}$. Note that $k_{1} =1$ because $\Op$ is a highest weight vector with weight 2.
To simplify notation, set $v^{j} := v^{p_{j}}$.
As before with highest weight vectors (see \eqref{ladder}),
we let $v^{j}_{l} = (1/l!)\Om^{l}.v^{j}$. 
Define
\leqn{xidef}{
\xi^{j} = \left\{
\begin{array}{ll}
i v^{j} + c(i v^{j}_{2k_{j}}) & 
\text{ if $c(v^{j}_{2k_{j}}) = -c(v^{j})$} \\ 
 v^{j} + c(v^{j}_{2k_{j}})  & \text{otherwise}
\end{array} 
\right.
}
for $j = 1,2,\ldots M$. Then straightforward calculation using \eqref{cmute} and 
\eqref{ladder} shows that $\Lo.\xi^{j} = 2k_{j} \xi^{j}$
and $\Op.\xi^{j} = 0$ for $j=1,2,\ldots M$. This implies that the
$\xi^{j}$ are all highest weight vectors of weight $2k_{j}$. Let
$V(\xi^{j})$ denote the irreducible submodule generated by $\xi^{j}$.
From \eqref{cinvar} and \eqref{xidef} it is clear that 
$\xi^{j} \subset V(2k_{j},v^{j})$ and hence $V(\xi^{j}) =  V(2k_{j},v^{j})$.
Thus the decomposition \eqref{gdecomp} shows that the sum
$\sum_{j=1}^{M} V(\xi^{j})$ is direct.

\emph{(iii)}: The relationship \eqref{xiconj} follows from
\eqref{cmute} , \eqref{xidef}, and \eqref{ladder}.

\emph{(iv)}: Because the numbers $2k_{1}, 2k_{2}, \ldots, 2k_{M}$
exhaust all the positive even weights and the sum 
$\sum_{j=1}^{M} V(\xi^{j})$ is direct, it follows from $\Sl{2}$-representation
theory that $\{\xi^{j} \; |\; j=1,2,\ldots M\}$ is a basis over $\mathbb{C}$ for $V_{2}$.
But $\{\mathbf{e}_{\alpha} \; | \; \alpha \in S_{\lambda} \}$ is
also a basis over $\mathbb{C}$ for $V_{2}$. Therefore we must
have $M = |S_{\lambda}|$.
\end{proof}

Define an $\mathbb{R}$-linear operator $A : \g \rightarrow \g$ by
\leqn{Adef}{
A = \half \ad(\Op) \circ \bigl(\,\ad(\Om) + \ad(\Op)\circ c\,\bigr) \; .
}

\begin{prop} \label{diag}\mnote{[diag]}
The $\mathbb{R}$-linear operator $A$ is symmetric with respect to
the inner product $\rip{\;\,}{\;}$, i.e.
$\rip{A(X)}{Y} = \rip{X}{A(Y)} \quad \forall \; X,Y \in \g $ .
\end{prop}

\begin{proof}
From \eqref{lam} and the properties \eqref{rip} of the inner product $\rip{\;\,}{\;}$,
it is not hard  to show that
$\rip{[\Op,[\Om,X]]}{Y}  = \rip{X}{[\Op,[\Om,Y]]}$ and
$\rip{[\Op,[\Op,c(X)]]}{Y} = \rip{X}{[\Op,[\Op,c(Y)]]}$ for every $X,Y \in \g$.
From the definition of $A$, it is then obvious that 
$\rip{A(X)}{Y} = \rip{A}{A(Y)}$ for every $X,Y \in \g$.
\end{proof}

\begin{lem}\label{AVp}\mnote{[AVp]}
\leqn{AV}{ 
A(V_{2}) \subset V_{2}
}
\end{lem}

\begin{proof}
It follows from $\Sl{2}$-representation theory that 
$\Omega_{\pm} . V_{n} \subset V_{n\pm 2}$. From \eqref{cmute}
it is clear that $c(V_{n})\subset V_{n}$. Thus 
$\Op . \Om . V_{2} \subset V_{2}$ and $\Op . \Op . c(V_{2}) \subset V_{2}$
which implies that $A(V_{2}) \subset V_{2}$.
\end{proof} 

This proposition shows that $A$ restricts to an operator on $V_{2}$. 
We denote this operator by 
\leqn{A2}{
A_{2} := A\restr{V_{2}} \; .
}

Label the integers $k_{j}$ from proposition  \ref{hwv} as 
follows
\begin{multline*}
1 = k_{J_{1}} = k_{J_{1}+1} = \cdots =  k_{J_{1}+m_{1}-1} 
<  k_{J_{2}} = k_{J_{2}+1} = \cdots =  k_{J_{2}+m_{2}-1} \\
< \cdots <  k_{J_{I}} = k_{J_{I}+1} = \cdots =  k_{J_{I}+m_{I}-1} \; ,
\end{multline*}
where
$J_{1} = 1$, $\;J_{l} + m_{l} = J_{l+1}$ for $l = 1,2,\ldots, I$ 
and $J_{I+1} = M-1$. Define 
\leqn{kf}{
\ke_{l} := k_{J_{l}} \quad l=1,2,\ldots, I \; .
}
The set $\{\xi^{j}_{k_{j}-1} \; | \; j= 1,2,\ldots M\}$ forms
a basis over $\mathbb{C}$ of $V_{2}$ by  proposition \ref{hwv} (iv). Therefore 
the set of vectors $\{X^{l}_{s},Y^{l}_{s} \;|\;
l=1,2,\ldots,I \; ; \;  s=0,1,\ldots, m_{l}-1 \}$ where
\leqn{Xls}{
X^{l}_{s} := \left\{ 
\begin{array}{l}
\xi^{J_{l}+s}_{\ke_{l}-1} \quad \text{ if $\ke_{l}$ is odd} \\
i\xi^{J_{l}+s}_{\ke_{l}-1} \quad  \text{ if $\ke_{l}$ is even}
\end{array}
\right. \AND
Y^{l}_{s}  := i X^{l}_{s} \; ,
}
forms a basis of $V_{2}$ over $\mathbb{R}$.
By proposition \ref{diag}, we
know that $A$ is symmetric and therefore diagonalizable. This forces
$A_{2}$ to also be diagonalizable.  
The next lemma shows that $\{X^{l}_{s},Y^{l}_{s} \;|\;
l=1,2,\ldots,I \; s=0,1,\ldots, m_{l}-1 \}$ is in fact an eigenbasis of $A_{2}$.

\begin{lem}\label{A2xp}\mnote{[A2xp]}
\leqn{A2x}{ 
A_{2}(X^{l}_{s})  = \ke_{l}(\ke_{l}+1) X^{l}_{s} \AND
A_{2}(Y^{l}_{s})  = 0 \text{\ for\ } l=1,2,\ldots,I \AND s=0,1,\ldots,m_{l}-1 \; .
}
\end{lem}

\begin{proof}
Using the formulas \eqref{ladder} and proposition \ref{hwv} (iii) it
is easy to show that \\
$A(\xi^{j}_{k_{j}-1})  = \half k_{j}(k_{j}+1)\left(1+(-1)^{k_{j}-1}\right)
\xi^{j}_{k_{j}-1} $ and $ A(i\xi^{j}_{k_{j}-1})  = 
\half k_{j}(k_{j}+1)\left(1+(-1)^{k_{j}}\right) i\xi^{j}_{k_{j}-1}$ 
for $j = 1,2,\ldots M$. The proposition then follows from
\eqref{kf} and  \ref{Xls}. 
\end{proof}

An immediate consequence of this lemma is that
$\text{spec}(A_{2}) = \{0\} \cup \{ \ke_{j}(\ke_{j}+1)  \; | \; 
j = 1,2,\ldots I \}$ and $m_{j}$ is the dimension of the eigenspace corresponding to 
the eigenvalue $\ke_{j}(\ke_{j}+1)$. Note that
$I$ is the number of distinct positive eigenvalues of $A_{2}$. 

Define  
\leqn{El}{
E^{l}_{0}  = \rspan{\{Y^{l}_{s}\;|\; s =0,1,\ldots, m_{l}-1\}} \; , \quad
E^{l}_{+}  = \rspan{\{X^{l}_{s}\;|\; s =0,1,\ldots, m_{l}-1\}} \; ,
}
and 
\leqn{E0}{
E_{0}  = \bigoplus_{l=1}^{I} E^{l}_{0} \; , \qquad 
E_{+}  = \bigoplus_{l=1}^{I} E^{l}_{+} \; .
}
Then $E_{0} = \ker\left(A_{2}\right)$ and 
$E^{l}_{+}$ is the eigenspace of $A_{2}$ corresponding to the eigenvalue $\ke_{l}(\ke_{l}+1)$
Moreover, using proposition \ref{hwv}  (iv), we see that
\leqn{V2}{
V_{2} = E_{0}\oplus E_{+} \; .
}

\begin{lem} \label{proja} \mnote{[proja]}
Suppose $X\in V_{2}$. Then 
$X\in \bigoplus_{q=1}^{l} E^{q}_{0}\oplus E^{q}_{+}$ if and only if
$\Op^{\ke_{l}} .\, X = 0$.
\end{lem}

\begin{proof}
From the formulas \eqref{ladder}, we get
\eqn{Op}{
\Op^{q-1} .\, \xi^{l}_{k_{l}-1} = 
\left\{
\begin{array}{ll}
0 & \text{if $ q > k_{l}$} \\
d(q,k_{l})\xi^{l}_{k_{l}-q} & \text{if $ q\leq k_{l}$}
\end{array}
\right. \; ,
}
where $ d(q,r) = (q+r)!/(r+1)!$.  This implies that
\leqn{Xrel}{
\Op^{l-1} .\,  X^{q}_{p} =
\left\{
\begin{array}{ll}
0 & \text{if $ l > \ke_{q}$} \\
\beta_{q}\, d(l,\ke_{q})\, \xi^{J_{q}+p}_{\ke_{q}-l} &
\text{if $ l\leq \ke_{q}$}
\end{array}
\right. \; ,
}
and 
\leqn{Yrel}{
\Op^{l-1} .\, Y^{q}_{p} =
\left\{
\begin{array}{ll}
0 & \text{if $ l > \ke_{q}$} \\ 
\overline{\beta}_{q}\, d(l,\ke_{q})\,i\, \xi^{J_{q}+p}_{\ke_{q}-l} & 
\text{if $ l\leq \ke_{q}$}
\end{array}
\right. \; ,
}
where 
\eqn{betas}{
\beta_{q} = \left\{ \begin{array}{ll}
1 & \text{ if $\ke_{q}$ is odd} \\
i & \text{ if $\ke_{q}$ is even}
\end{array} \right. \; .
}
Suppose $X\in V_{2} = \bigoplus_{q=1}^{I} E^{q}_{0}\oplus E^{q}_{+}$. Then
there exists real constants $a_{q p}$ and $b_{q p}$ such that
\leqn{Xproj}{
X = \sum^{I}_{q=1}\sum^{m_{q}-1}_{p=0} \left( a_{q p}\, Y^{q}_{p} +
b_{q p} \, X^{q}_{p} \right) \; .
}
Suppose $\Op^{\ke_{l}} . \, X = 0$. Then \eqref{Xrel}, \eqref{Yrel},
and \eqref{Xproj} imply that
\eqn{sums}{
\sum^{I}_{q=l+1} \sum^{m_{q}-1}_{p=0} \left(
a_{q p}\,\overline{\beta}_{q}\,d(\ke_{l}+1,\ke_{q})\, i\, 
\xi^{J_{q}+p}_{\ke_{q}-\ke_{l}-1} +
b_{q p}\,\beta_{q}\,d(\ke_{l}+1,\ke_{q})\, 
\xi^{J_{q}+p}_{\ke_{q}-\ke_{l}-1} \right) = 0 \; .
}
But the set of vectors
\eqn{vecs}{
\left\{ \overline{\beta}_{q} \, i\,
\xi^{J_{q}+p}_{\ke_{q}-\ke_{l}-1}, \beta_{q}\,\xi^{J_{q}+p}_{\ke_{q}-\ke_{l}-1}
\; | \; q=l+1,l+2,\ldots I \; , \; p=0,1,\ldots, m_{q}-1 \; \right\}
}
is linearly independent over $\mathbb{R}$. Therefore 
$ X = \sum_{q=1}^{l} \sum_{p=0}^{m_{q}+1}
( a_{q p} Y^{q}_{p} + b_{q p} X^{q}_{p})$ which implies that
$ X \in \bigoplus_{q=1}^{l} E^{q}_{0} \oplus E^{q}_{+}$.

Conversely, suppose $ X \in \bigoplus_{q=1}^{l} E^{q}_{0} \oplus E^{q}_{+}$.
The $X$ can be written in the form \eqref{Xproj} and it is easy using
\eqref{Xrel} and \eqref{Yrel} to verify that  $\Op^{\ke_{l}} . \, X = 0$.
\end{proof}

\begin{lem} \label{projb}\mnote{[projb]}
Suppose $X\in V_{2}$. Then
$X\in \bigoplus_{q=1}^{l} E^{q}_{0}\oplus E^{q}_{+}$ if and only if
$\Op^{\ke_{l}+2} .\, c(X) = 0$.
\end{lem}

\begin{proof}
Proved in a similar fashion as lemma \ref{proja}.
\end{proof}

\begin{lem} \label{tildeprop}\mnote{[tildeprop]}
Let $\;\; \widetilde{\;} : \mathbb{Z}_{\geq -1}  \rightarrow \{1,2,\ldots,I\}$ be
the map defined by
\eqn{tildemap}{
\text{ $\widetilde{-1} = \tilde{0} = 1$ and  $\tilde{s} = \max\{ \;l\;|\;\ke_{l} \leq s 
\; \}$ if $s > 0$. }
}
Then
\begin{itemize}
\item[(i)] \hspace{1cm} $\ke_{\tilde{s}} \leq s$ for every 
$s \in \mathbb{Z}_{\geq 0}.$
\item[(ii)] \hspace{1cm} $\ke_{\tilde{s}} \leq s < \ke_{\tilde{s}+1}$
for every $s \in \{0,1,\ldots \ke_{I}-1\}$.
\end{itemize}
\end{lem}

\begin{proof}
\emph{(i)}  This is obvious from the definition of $\;\widetilde{\;}$ . 

\emph{(ii)} From part (i), $\ke_{\tilde{s}} \leq s$. So suppose
$\ke_{\tilde{s}+1} \leq s$. Then
from the definition of  $\;\widetilde{\;}$ it is clear 
$\ke_{\tilde{s}+1} \leq \ke_{\tilde{s}}$. But because 
$\ke_{1} < \ke_{2} < \cdots
< \ke_{I}$, it follows that $\tilde{s}+1 \leq s$ which is a contradiction.
Thus $\ke_{\tilde{s}+1} > s$ and we are done.
\end{proof}

\begin{lem} \label{projc}\mnote{projc}
If $X\in V_{2}$, $\ke_{\tilde{p}} + s < \ke_{\tilde{p}+1}$ 
$(s\geq 0)$, and $\Op^{\ke_{\tilde{p}}+s} . \, X = 0$, then
$\Op^{\ke_{\tilde{p}}}  . \, X = 0$.
\end{lem}

\begin{proof}
Assume $s > 0$, otherwise we are done. Because $X\in V_{2}$, we
have $\Op^{\ke_{\tilde{p}}+s-1} . \, X \in V_{2(\ke_{\tilde{p}}+s)}$.
By assumption $\Op^{\ke_{\tilde{p}}+s} . \, X = 0$, so
\eqn{OpV2}{
 \Op^{\ke_{\tilde{p}}+s-1} . \, X \in  V_{2(\ke_{\tilde{p}}+s)} \cap
\ker(\ad(\Op)) \; .
}
But, if $n \in \mathbb{Z}_{>0}$, then
\eqn{V2n}{
V_{2n} \cap \ker(\ad(\Op)) \neq \{0\} \Longleftrightarrow 
n \in \{ \ke_{1}, \ke_{2}, \ldots, \ke_{I}\} \; ,
}
because otherwise $\g$ would contain an irreducible  
$\cspan{\{\Lo,\Op,\Om\}}$-submodule
with weight $2n \in  \mathbb{Z}_{>0}\backslash  
\{ 2\ke_{1},2\ke_{2}, \ldots, 2\ke_{I}\}$. This is impossible as
the set $\{ 2\ke_{1}, 2\ke_{2}, \ldots, 2\ke_{I}\}$ exhausts all
the positive even weights of the irreducible
$\cspan{\{\Lo,\Op,\Om\}}$-submodules in $\g$. Therefore
$\Op^{\ke_{\tilde{p}}+s-1} . \, X = 0 $ as $\ke_{\tilde{p}} < \ke_{\tilde{p}}+s < \ke_{\tilde{p}+1}$
implies that $(\ke_{\tilde{p}}+s)$ is not in $\{ \ke_{1},\ke_{2}, \ldots, \ke_{I}\}$. Repeat the above
argument with $s'=s-1$ to arrive at  $\Op^{\ke_{\tilde{p}}+s'-1} . \, X = 
 \Op^{\ke_{\tilde{p}}+s-2} . \, X = 0$. Continuing in this manner, we
find $\Op^{\ke_{\tilde{p}}} . \, X = 0$.
\end{proof} 

The next theorem is the key result needed to prove that
the EYM equations can be put into a form where theorem \ref{BFM}
applies in a neighborhood of the origin $r=0$.

\begin{thm} \label{Liebfma}\mnote{[Liebfma]}
Suppose $p \in \{ 1,2,\ldots,\ke_{I}-1\}$ and $Z_{0},Z_{1},\ldots,Z_{p+1}
\in V_{2}$ is a sequence of vectors that satisfy
$Z_{0} \in E^{1}_{0}\oplus E^{1}_{+}$ and 
$Z_{n+1} \in \bigoplus_{q=1}^{\tilde{n}} E^{q}_{0}\oplus E^{q}_{+}$ for
$n =0,1,\ldots p$.  Then for every $j \in \{1,2,\ldots,p+1\}$ , 
$s\in \{0,1,2,\ldots,j\}$
\begin{itemize}
\item[(i)] \hspace{1cm} $[[c(Z_{j-s}),Z_{s}],Z_{p+2-j}] \in 
\bigoplus_{q=1}^{\tilde{p}} E^{q}_{0}\oplus E^{q}_{+}$
\item[(ii)] \hspace{1cm} $[[c(Z_{p+2-j}),Z_{j-s}],Z_{s}] \in 
\bigoplus_{q=1}^{\tilde{p}} E^{q}_{0}\oplus E^{q}_{+}$
\end{itemize}
\end{thm}

\begin{proof}
\emph{(i)} Suppose $Z_{0},Z_{1},\ldots,Z_{p+1} \in V_{2}$ is a sequence
satisfying $Z_{0} \in E^{1}_{0}\oplus E^{1}_{+}$ and       
$Z_{n+1} \in \bigoplus_{q=1}^{\tilde{n}} E^{q}_{0}\oplus E^{q}_{+}$ for
$n =0,1,\ldots p$. Then 
\leqn{Liebfma1}{
\Op^{\ke_{(n-1)\widetilde{\;}}} . \, Z_{n} = 
\Op^{\ke_{(n-1)\widetilde{\;}}+2} . \, c(Z_{n}) = 0 
}
for $n = 0,1,2,\ldots,p+1$ by lemmas \ref{proja} and
\ref{projb}. Now, if $j \in \{1,2,\ldots,p+1\}$ and
$s\in\{0,1,\ldots,j\}$, then
\eqn{Lfm}{ 
\Op^{p}. [[c(Z_{j-s}),Z_{s}],Z_{p+2-j}] = \sum_{l=0}^{p} \sum_{m=0}^{l}
\binom{p}{l}\binom{l}{m} a_{psjlm}
}
where
\eqn{aps}{
a_{psjlm} =  [[\Op^{m} .\, c(Z_{j-s}), \Op^{l-m} . \, Z_{s}],
\Op^{p-l}. \, Z_{p+2-j}] \; .
}
Applying \eqref{Liebfma1} yields
$a_{psjlm} = 0$ if $m-2\geq \ke_{(j-s-1)\widetilde{\;}}$ or
$l-m \geq \ke_{(s-1)\widetilde{\;}}$ or $p-l\geq \ke_{(p+1-j)\widetilde{\;}}$.
But because of lemma \ref{tildeprop} (i), this implies
that $a_{psjlm} = 0$ if $m-2 \geq j-s-1$ or $l-m\geq s-1$ or 
$p-l \geq p+1-j$. It follows that  $a_{psjlm} = 0$ unless $l$ and $m$
satisfy $j-1 < l < m+s-1 < j$ which is impossible. Therefore 
$a_{psjlm} = 0$ for all $l$ and $m$. Thus 
$\Op^{p}. [[c(Z_{j-s}),Z_{s}],Z_{p+2-j}] = 0$. But then it follows
from lemmas \ref{tildeprop} (ii) and \ref{projc} that
$\Op^{\tilde{p}}. [[c(Z_{j-s}),Z_{s}],Z_{p+2-j}] = 0$ and hence
$[[c(Z_{j-s}),Z_{s}],Z_{p+2-j}] \in
\bigoplus_{q=1}^{\tilde{p}} E^{q}_{0}\oplus E^{q}_{+}$ by lemma
\ref{proja}.

\emph{(ii)} It follows from similar arguments that 
$[[c(Z_{p+2-j}),Z_{j-s}],Z_{s}] \in \bigoplus_{q=1}^{\tilde{p}} 
E^{q}_{0}\oplus E^{q}_{+}$.
\end{proof}

It is worthwhile to note that all the the above results did not 
depend on $\Lo$ being regular.
However, for what follows we will need $\Lo$ to be regular.

\begin{prop} \label{Abelian}\mnote{[Abelian]}
Suppose $\Lo$ is regular. Then $\cspan{\{\xi^{1},\xi^{2},\ldots,\xi^{M}\}}$
is an Abelian subalgebra of $\g_{\lambda}$ and hence also an Abelian
subalgebra of $\g$.
\end{prop}

\begin{proof}
From the definition of $\g_{\lambda}$, it follows that
$\cspan{\{\Lo,\Op,\Om\}} \subset \g_{\lambda}$ and
$V_{2} \subset \g_{\lambda}$. But by proposition \ref{hwv}
$V_{2} = \cspan{\{\xi^{1}_{k_{1}-1},\xi^{2}_{k_{2}-1},\xi^{M}_{k_{M}-1}\}}$,
and hence
\eqn{kl1}{
\frac{(k_{l}+1)!}{(2k_{l})!}\Op^{k_{l}-1} . \, \xi^{l}_{k_{l}-1}
= \xi^{l} \in \g_{\lambda}
}
for $i = 1,2,\ldots, M$. 
Therefore $\cspan{\{\xi^{1},\xi^{2},\ldots,\xi^{M}\}}
\subset \g_{\lambda}$. The $\xi^{j}$ are highest weight vectors,
consequently 
\leqn{Abelian1}{
\cspan{\{\xi^{1},\xi^{2},\ldots,\xi^{M}\}} \subset \g_{\lambda}^{\Op} 
}
where $\g_{\lambda}^{\Op} = \{\; X\in \g_{\lambda} \; | \; [\Op,X] = 0 \; \}$.
Define $V_{\lambda,n} := \{\; X\in \g_{\lambda} \; | \; \Lo .\, X = n X \;\}$.
By theorem \ref{pisys}, $S_{\lambda}$ is a base
a system of roots of $\g_{\lambda}$ and $\alpha(\Lo) = 2$ for every
$\alpha \in S_{\lambda}$ and hence it follows that
$V_{\lambda,{2}} = V_{2}$. Using $\Sl{2}$-representation theory, it
is not hard to show that $\dim_{\mathbb{C}} \g_{\lambda}^{\Op}
= \dim_{\mathbb{C}} V_{\lambda,2}$. But $\dim_{\mathbb{C}} V_{2} = 
|S_{\lambda}|$, and therefore 
$\dim_{\mathbb{C}} \g_{\lambda}^{\Op} = |S_{\lambda}|$. By
proposition \ref{hwv}, $|S_{\lambda}| = M$ and
hence we get from \eqref{Abelian1} that 
\leqn{Abelian2}{
\cspan{\{\xi^{1},\xi^{2},\ldots,\xi^{M}\}} = \g_{\lambda}^{\Op} \; .
}
Theorem \ref{pisys} proved that  
$|S_{\lambda}| = \dim_{\mathbb{C}}\h_{\lambda}$ which in turn gives
, via the above result, 
$\dim_{\mathbb{C}}\g_{\lambda}^{\Op} = \dim_{\mathbb{C}}\h_{\lambda} \; .$
Applying lemma 2.1.15 of \cite{k6494} then shows that
\leqn{Abelian3}{
\dim_{\mathbb{C}} \g_{\lambda}^{\Op} = \min \{\; \dim_{\mathbb{C}}
\g_{\lambda}^{X} \; | \; X \in  \g_{\lambda} \; \} \; .
}
We can identify $\g_{\lambda}$ with the dual $\g_{\lambda}^{*}$ using the 
form $(\; , \; )$, i.e.
\eqn{iota}{
\iota :  \g_{\lambda,0} \rightarrow \g_{\lambda,0}^{*} \;\; ; \;\;
\iota(X)(\cdot) = (X,\cdot) \; .
}
So if $f\in \g_{\lambda}^{*}$ and we define 
$\g_{\lambda}^{f} = \{\;X\in \g_{\lambda} \; |\; \ad^{*}_{X}(f)=0\}$,
then it can be shown that 
\leqn{Abelian4}{
\g_{\lambda}^{\iota(X)} = 
\g_{\lambda}^{X} \quad \forall \; X\in\g \; .
} 
Let $G_{\lambda}$ be a connected
complex semisimple Lie group with Lie algebra $\g_{\lambda}$. Then for  $f\in \g_{\lambda}^{*}$,
$\g_{\lambda}^{f}$ is the Lie algebra of coadjoint isotropy group 
$G_{\lambda,f} = \{ a \in G_{\lambda} \; | \text{Ad}^{*}_{a}(f) = f \; \}$.
But then  \eqref{Abelian2}, \eqref{Abelian3}, \eqref{Abelian4} and 
a straightforward generalization of theorem 9.3.10 in \cite{k6081} to complex
Lie groups imply that $\cspan{\{\xi^{1},\xi^{2},\ldots,\xi^{M}\}}$ is
an Abelian subalgebra.
\end{proof}

The next theorem is the key result needed to prove that
the EYM equations can be put into a form where theorem \ref{BFM}
applies in a neighborhood $r=\infty$. Although this theorem
looks very similar to theorem \ref{Liebfma}, it is more
difficult to prove. Similar arguments as in 
theorem \ref{Liebfma} are employed, but these only go
part of the way. Proposition \ref{Abelian} is needed 
to complete the proof. 

\begin{thm} \label{Liebfmb}\mnote{[Liebfmb]}
Assume that $\Lo$ is regular. Suppose $p\in\{0,1,2,\ldots,\ke_{I}\}$
and $Z_{0},Z_{1},\ldots, Z_{p} \in V_{2}$ is a sequence of vectors
that satisfy $Z_{n} \in \bigoplus_{q=1}^{\tilde{n}} E^{q}_{0}\oplus E^{q}_{+}$
for $n = 0,1,2\ldots p$. Then for every $j\in \{1,2,\ldots,p\}$ ,
$s\in\{0,1,\ldots,j\}$
\begin{itemize}
\item[(i)] \hspace{1cm} $[[c(Z_{j-s}),Z_{s}],Z_{p+1-j}] \in
\bigoplus_{q=1}^{\tilde{p}} E^{q}_{0}\oplus E^{q}_{+}$
\item[(ii)] \hspace{1cm} $[[c(Z_{p+1-j}),Z_{j-s}],Z_{s}] \in
\bigoplus_{q=1}^{\tilde{p}} E^{q}_{0}\oplus E^{q}_{+}$
\end{itemize}
\end{thm}
\begin{proof}
\emph{(i)} Suppose $Z_{0},Z_{1},\ldots,Z_{p} \in V_{2}$ is a sequence
satisfying $Z_{n} \in \bigoplus_{q=1}^{\tilde{n}} E^{q}_{0}\oplus E^{q}_{+}$ for
$n =0,1,\ldots p$. Then 
\leqn{Liebfmb1}{
\Op^{ \ke_{\tilde{n}} } . \, Z_{n} =
\Op^{ \ke_{\tilde{n}}+2 } . \, c(Z_{n}) = 0
}
for $n = 0,1,2,\ldots,p$ by lemmas \ref{proja} and
\ref{projb}. Suppose $j \in \{1,2,\ldots,p\}$ and
$s\in\{0,1,\ldots,j\}$. Then
\leqn{Liebfmb2}{
\Op^{p}. [[c(Z_{j-s}),Z_{s}],Z_{p+1-j}] = \sum_{l=0}^{p} \sum_{m=0}^{l}
\binom{p}{l}\binom{l}{m} a_{psjlm}
}
where
\eqn{apsj}{
a_{psjlm} =  [[\Op^{m} .\, c(Z_{j-s}), \Op^{l-m} . \, Z_{s}],
\Op^{p-l}. \, Z_{p+1-j}] \; .
}
Applying \eqref{Liebfmb1} yields
$a_{psjlm} = 0$ if $m-2\geq \ke_{(j-s)\widetilde{\;}}$ or
$l-m \geq \ke_{\tilde{s}}$ or $p-l\geq \ke_{(p+1-j)\widetilde{\;}}$.
But because of lemma \ref{tildeprop} (i), this implies
that $a_{psjlm} = 0$ if $m-2 \geq j-s$ or $l-m\geq s$ or
$p-l \geq p+1-j$. It follows that  $a_{psjlm} = 0$ unless $l$ and $m$
satisfy $j-1 < l < m+s < j+2$ which implies that $l=j$ and
$m+s=j+1$. Thus the sum \eqref{Liebfmb2} reduces to
\eqn{Opp}{
\Op^{p}. [[c(Z_{j-s}),Z_{s}],Z_{p+1-j}] = \binom{p}{j}\binom{j}{j+1-s}
[[X_{1},X_{2}],X_{3}]
}
where $X_{1} = \Op^{j-s+1} .\, c(Z_{j-s})$, $X_{2} = \Op^{s-1} . \, Z_{s}$ , and
$X_{3} = \Op^{p-j}. \, Z_{p+1-j}$. Applying \eqref{Liebfmb1} then shows that
$\Op .\, X_{a} = 0$ for $a=1,2,3$. Because the $X_{a}$ have even weights, 
\eqn{X1}{
X_{1}, X_{2}, X_{3} \in \cspan{\{\xi^{1},\xi^{2},\ldots,\xi^{M}\}} \;.
}
But $ \cspan{\{\xi^{1},\xi^{2},\ldots,\xi^{M}\}}$ is an Abelian subalgebra by proposition
\ref{Abelian}, so $[[X_{1},X_{2}],X_{3}]=0$ which implies that 
$\Op^{p}. [[c(Z_{j-s}),Z_{s}],Z_{p+1-j}] = 0$. We then get via lemma \ref{projc} 
that \\
$\Op^{\ke_{\tilde{p}}} . [[c(Z_{j-s}), Z_{s}], Z_{p+1-j}] = 0$ and hence
$[[c(Z_{j-s}),Z_{s}],Z_{p+1-j}] \in \bigoplus_{q=1}^{\tilde{p}} E^{q}_{0}\oplus E^{q}_{+}$
by lemma \ref{proja}.

\emph{(ii)} The proof that $[[c(Z_{p+1-j}),Z_{j-s}],Z_{s}] \in
\bigoplus_{q=1}^{\tilde{p}} E^{q}_{0}\oplus E^{q}_{+}$ is similar to part \emph{(i)}.
\end{proof}

\begin{prop} \label{Eplus} \mnote{[Eplus]}
If $\Op \in \sum_{\alpha\in S_{\lambda}} \mathbb{R}\mathbf{e}_{\alpha}$ and $\Lo$ is
regular, then $E_{+} =  \sum_{\alpha\in S_{\lambda}} \mathbb{R}\mathbf{e}_{\alpha}$.
\end{prop}

\begin{proof}
Introduce a basis $\{\; Z_{j}\; | \;  1\leq j \leq M \; \}$ over $\mathbb{R}$ for
$E_{+}$ by defining
\eqn{Zj}{
Z_{j} = \left\{
\begin{array}{ll}
\xi^{j}_{k_{j}-1} & \text{if $k_{j}$ is odd} \\
i\, \xi^{j}_{k_{j}-1} & \text{if $k_{j}$ is even}
\end{array}
\right. \quad 1\leq j \leq M \; .
}
Equations \eqref{ladder} and proposition \ref{hwv} (iii) can be used to
show that
\leqn{Eplus1}{
\Op . \, c(Z_{j}) = \Om . \, Z_{j} \quad 1\leq j \leq M \; . 
}
By assumption $\Op = \sum_{\alpha\in S_{\lambda}} w_{\alpha} \mathbf{e}_{\alpha}$ for
some set of constants $w_{\alpha} \in \mathbb{R}$. Because $c(\Op) = -\Om$ and 
$c( \mathbf{e}_{\alpha}) = -  \mathbf{e}_{-\alpha}$,
 $\Om = \sum_{\alpha\in S_{\lambda}} w_{\alpha} \mathbf{e}_{-\alpha}$. Since
$Z_{j} \in V_{2}$,
$Z_{j} = \sum_{\alpha\in S_{\lambda}} a_{j\alpha}  \mathbf{e}_{\alpha}$ for
some set of constants $a_{j\alpha} \in \mathbb{C}$. So then 
$c(Z_{j}) = -\sum_{\alpha\in S_{\lambda}} \overline{a}_{j\alpha}  \mathbf{e}_{-\alpha}$.
Now, since $\Lo$ is regular, equation \eqref{eab} holds. 
Therefore
\leqn{Eplus2}{
\Om . \, Z_{j} = \sum_{\alpha\in S_{\lambda}} w_{\alpha} a_{j\alpha} 
[ \mathbf{e}_{-\alpha}, \mathbf{e}_{\alpha}] 
=  \sum_{\alpha\in S_{\lambda}} - w_{\alpha} a_{j\alpha} \mathbf{h}_{\alpha} \; , 
}
while
\leqn{Eplus3}{
\Op . \, c(Z_{j}) = \sum_{\alpha\in S_{\lambda}} -w_{\alpha} \overline{a}_{j\alpha} 
[ \mathbf{e}_{\alpha}, \mathbf{e}_{-\alpha}] 
=  \sum_{\alpha\in S_{\lambda}} - w_{\alpha} \overline{a}_{j\alpha} \mathbf{h}_{\alpha} \; .
}
The three results \eqref{Eplus1}, \eqref{Eplus2}, and \eqref{Eplus3} then
yield
\eqn{sumal}{
\sum_{\alpha\in S_{\lambda}} w_{\alpha} ( a_{j\alpha} - \overline{a}_{j\alpha}) \mathbf{h}_{\alpha}
= 0 \;.
}
Since $\Lo$ is regular, it follows that $w_{\alpha} \neq 0$ for all $\alpha \in 
S_{\lambda}$ and the set $\{\;\mathbf{h}_{\alpha}\;|\; \alpha \in S_{\lambda}\;\}$ is
linearly independent \cite{k4295}. Thus  $a_{j\alpha} - \overline{a}_{j\alpha} = 0$ for all
$\alpha \in S_{\lambda}$ and $ j = 1,2,\ldots,M$.  So $Z_{j} \in \sum_{\alpha\in S_{\lambda}} 
\mathbb{R}\mathbf{e}_{\alpha}$ for $j=1,2,\ldots, M$ which implies that
$E_{+} \subset \sum_{\alpha\in S_{\lambda}} \mathbb{R}\mathbf{e}_{\alpha}$.
However, $\dim_{\mathbb{R}} E_{+} = \dim_{\mathbb{R}} (\sum_{\alpha\in S_{\lambda}} 
\mathbb{R}\mathbf{e}_{\alpha}) = |S_{\lambda}|$ and therefore $E_{+} =
\sum_{\alpha\in S_{\lambda}} \mathbb{R}\mathbf{e}_{\alpha}$.
\end{proof}

Suppose  $\Lo$ is regular and $\Op = \sum_{\alpha\in S_{\lambda}} w_{\alpha} \mathbf{e}_{\alpha}$
where $w_{\alpha} \in \mathbb{R}$ for every $\alpha \in S_{\lambda}$. Then using 
\eqref{chev}, \eqref{eab}, and the fact that $c(\mathbf{e}_{\alpha}) = -\mathbf{e}_{-\alpha}$,
it is not difficult to show that
\eqn{A2matrix1}{
A_{2}(\mathbf{e}_{\alpha}) =  \sum_{\beta \in S_{\lambda}} w_{\beta} \hbr{\beta}{\alpha} w_{\alpha}
\mathbf{e}_{\beta}  \; .
}
This result along with \eqref{V2} and proposition \ref{Eplus} shows that
$\{\: \mathbf{e}_{\alpha}\:|\: \alpha \in S_{\lambda}\: \}$ can be completed
to a basis over $\mathbb{R}$ of $V_{2}$ so that the matrix of $A_{2}$ with respect
to this basis takes the form
\leqn{A2matrix2}{
[A_{2}] = \left(
\begin{array}{cc}
 0 & 0 \\
 0 & [A_{\alpha \beta}]
\end{array}
\right) \; ,
}
with  
\leqn{A2matrix3}{
A_{\alpha \beta} = w_{\alpha} \hbr{\alpha}{\beta} w_{\beta} \; . 
}
 
\sect{BFM2}{Local Uniqueness and Existence Proofs}

In this section we present the proofs of theorems
\ref{exist0}, \ref{existInf}, and \ref{existHor}.
The proof of theorem \ref{existHor} is the easiest
and does not depend on the results of the section \ref{BFM1}.

Define
\eqn{evalues}{
        \Ec := \{\;\ke_{j} \; | \; j=1,2,\ldots I \} \; 
}
with the $\ke_{j}$ defined in $\eqref{kf}$
and let
\eqn{eproj}{
\pr{q} : E_{+} \rightarrow E^{q}_{+} \quad q = 1,2,\ldots I
}
denote the projection operators between the spaces defined in 
\eqref{El} and \eqref{E0}. If $a \in \mathbb{R}$, we will use $I_{\epsilon}(a)$
to denote an interval of radius $\epsilon$ about $a$, i.e.
\eqn{Rinterval}{
I_{\epsilon}(a) = (a-\epsilon,a+\epsilon) \; .
}

From proposition \ref{Eplus} and \eqref{nfeq4}, we know that the the 
solution $\Lp(r)$ to equation \eqref{feq4} is, up to a gauge transformation,
completely characterized by the condition
\leqn{realLp}{
      \Lp(r) \in E_{+} \quad \forall \; r \; .
}
As discussed previously, if we can solve the two EYM equations
\eqref{feq1} and \eqref{feq3} for the variables $\{\Lp(r),m(r)\}$ the
the remaining equation \eqref{feq2} can be integrated to yield
$S$. Consequently, we are only interested in the equations
\eqref{feq1} and \eqref{feq3}. 

\begin{proof}[Proof of theorem \ref{exist0}] 
The proof of this theorem involves finding a change of variables
to put the system of differential equations \eqref{feq1} and \eqref{feq3} into 
a form where theorem \ref{BFM} applies in a neighborhood of $r=0$. 

Since $\Lp$ satisfies \eqref{realLp}, we can introduce new variables
$\{u_{s+1}(r) \; | \; s\in \Ec\;\}$ that satisfy
\leqn{origbfm1}{
\Lp(r) = \Op  + \sum_{s\in \Ec} u_{s+1}(r)\, r^{s+1}
}
where $\Op = \Lp(0)$ and $u_{s+1}(r) \in E^{\tilde{s}}_{+}$ for all $r$ and $s \in \Ec$.
Because $E_{+} = \bigoplus_{q=1}^{I}  E^{q}_{+}$, it is obvious that this transformation
is invertible. Define
\eqn{origbfm2}{
\chi_{s+1} = \left\{
\begin{array}{ll}
1 & \text{ if $s\in \Ec$} \\
0 & \text{ otherwise}
\end{array}
\right. \; ,
}
Then we can write $\Lp(r) = \Op  + \sum_{k=0}^{\infty}\chi_{k} u_{k}(r)\,r^{k}$.
Substituting this in \eqref{vardefs5} shows that there exists an integer
$N_{1}$ such that
\eqn{origbfm3}{
\Fc = -\sum_{k\in \Ec} A_{2}(u_{k+1})\,r^{k+1} + \sum_{k=2}^{N_{1}} f_{k} r^{k}  \; ,
}
where 
\alin{origbfm4}{
f_{k} = & \half \sum_{j=2}^{k-2} \Bigl\{ [[\Op,c(\chi_{j} u_{j})] + 
        [\Om,\chi_{j} u_{j}], \chi_{k-j} u_{k-j}]   \nonumber \\
      &  + [[\chi_{j} u_{j},c(\chi_{k-j} u_{k-j})], \Op] 
        + \sum_{s=2}^{j-2}  [[\chi_{s} u_{s},c(\chi_{j-s} u_{j-s})], \chi_{k-j} u_{k-j} ]
        \Bigr\} \; .
}
But $A_{2}(u_{k+1}) = k(k+1) u_{k+1}$ for
every $k \in \Ec$ by lemma \ref{A2xp} and hence
\leqn{origbfm5}{
\Fc = -\sum_{k\in \Ec} k(k+1) u_{k+1}\,r^{k+1} + \sum_{k=2}^{N_{1}} f_{k} r^{k}  \; .
}
Define
\leqn{origbfm6}{
v_{s+1} = u_{s+1}' \quad \forall \; s \in \Ec \; .  
}
Using \eqref{origbfm1}, \eqref{origbfm5} and \eqref{origbfm6},
the EYM equation \eqref{feq3} can be written as 
\lalign{origbfm7}{
r\sum_{k\in\Ec} v_{k+1}'\, r^{k+1} = & -2 \sum_{k\in\Ec} (k+1) v_{k+1}\, r^{k+1}
+ \sum_{k\in\Ec} \frac{k(k+1)}{r} \left(\frac{1}{N}-1\right) u_{k+1} \, r^{k+1} \nonumber \\
& -\frac{2}{r N}\left(m-\frac{1}{r} P\right)  \sum_{k\in\Ec} \left( 
v_{k+1} \, r^{k+1} + (k+1) u_{k+1} r^{k} \right) - \frac{1}{N}  \label{origbfm7.2}
\sum_{k=4}^{N_{1}} f_{k} \, r^{k-1} \;.
} 
Applying the projections $\pr{\tilde{k}}$ for every $ k \in \Ec$ to equation 
\eqref{origbfm7.2} yields
\lalign{origbfm8}{
r v_{k+1}'= & -2 (k+1) v_{k+1} - \frac{2}{r N}\left(m-\frac{1}{r} P\right)
v_{k+1}  + \frac{k(k+1)}{r} \left(\frac{1}{N}-1\right) u_{k+1} \nonumber \\
&  -\frac{2}{r^{2} N}\left(m-\frac{1}{r} P\right) (k+1) u_{k+1} r^{k} - 
\frac{1}{r^{k+1}N} \sum_{s=2}^{N_{1}-2} \pr{\tilde{k}}\left(f_{s+2}\right) \, r^{s+1} \quad
 \forall \; k\in \Ec \; .\label{origbfm8.2}
} 
The last term in \eqref{origbfm8.2} is the main obstruction to putting the
equation into a form where theorem \ref{BFM} applies.  It seems to contain
terms of order $r^{-s}$ $\:(s > 0)$ . However, as we shall now see, the
results of section \ref{BFM1} can be used to show that
\leqn{origbfm8a}{
\frac{1}{r^{k+1}N} \sum_{s=0}^{N_{1}-2} \pr{\tilde{k}}\left(f_{s+2}\right) \, r^{s+1}
= \frac{1}{N} \sum_{s=k}^{N_{1}-2} \pr{\tilde{k}}\left(f_{s+2}\right) \, r^{s-k} \; .
}
Namely, by using proposition \ref{Eplus}, we can show that $f_{k} \in E_{+}$ for
all $k$. From the definition of the $u_{s+1}$ it is clear that 
$\chi_{s+1} u_{s+1} \in \bigoplus^{\tilde{s}}_{q=1} E_{+}^{q}$
for $0 \leq s \leq \ke_{I}$, and so it follows from theorem \ref{Liebfma}
by letting $Z_{0} = \Op$ and $Z_{k+1} = \chi_{k+1} u_{k+1}$ for $k\geq 0$
that $f_{s+2} \in \bigoplus_{q=1}^{\tilde{s}} E_{+}^{q}$. Consequently, 
for every $k\in \Ec$
\eqn{origbfm9}{
\pr{\tilde{k}}\left( f_{s+2}\right) = 0 \quad \text{ if $s < k$} \; ,
}
because $k\in \Ec$ implies that $k = \ke_{\tilde{k}}$ and hence 
it follows for $s < k=\ke_{\tilde{k}}$ that $\tilde{s} < \tilde{k}$.
This proves \eqref{origbfm8a}.
Therefore we can rewrite \eqref{origbfm8.2} as
\lalign{origbfm10}{
r v_{k+1}'= & -2 (k+1) v_{k+1} - \frac{2}{r N}\left(m-\frac{1}{r} P\right)
v_{k+1}  + \frac{k(k+1)}{r} \left(\frac{1}{N}-1\right) u_{k+1} \nonumber \\
&  -\frac{2}{r^{2} N}\left(m-\frac{1}{r} P\right) (k+1) u_{k+1} -
\frac{r}{N} \sum_{s=k}^{N_{1}-1} \pr{\tilde{k}}\left(f_{s+3}\right) \, r^{s-k}  \nonumber \\
& + \left(1-\frac{1}{N}\right) \pr{\tilde{k}} \left( f_{k+2}\right) - \pr{\tilde{k}} \left( f_{k+2} \right)
\quad
 \forall \; k\in \Ec \; .\label{origbfm10.2}
}

Using the properties \eqref{rip} of $\rip{\;\,}{\;}$ and the fact that
$A_{2}(u_{2}) = 2 u_{2}$, it can be shown that there exists analytic functions
\eqn{origbfm11}{
\hat{P}  : E_{+} \times \mathbb{R} \rightarrow \mathbb{R} \AND
\hat{G}  : E_{+} \times E_{+} \times \mathbb{R} \rightarrow \mathbb{R} \; ,
}
such that
\leqn{origbfm12}{
P  = r^{4}\norm{u_{2}}^{2} + r^{5}\hat{P}(u,r) \AND
G  = r^{2} 2\norm{u_{2}}^{2} + r^{3}\hat{G}(u,v,r) \; ,
}
where $ u = \sum_{s\in \Ec} u_{s+1} $, $v = \sum_{s\in \Ec} v_{s+1}$,
and $\norm{\;}$ is defined by \eqref{norm}.
Introduce a new ``mass'' variable $\mu$ by
\leqn{origbfm13}{
\mu = \frac{1}{r^{3}}(m-r^{3} \norm{u_{2}}^{2}) \; .
}
Recall that $\ke_{1}=1$, so $1$ is always in $\Ec$ and hence
$u_{2}$ is always defined.  We can then write the EYM equation \eqref{feq1} as
\lalign{origbfm14}{
r \mu' = -3 \mu + r&\left\{  \hat{P}(u,r) + \hat{G}(u,v,r) - 
        2 \rip{u_{2}}{v_{2}} \right. \nonumber \\ 
        & \left. - 2\, r\, (\mu + \norm{u_{2}}^{2}) 
        ( 2 \norm{u_{2}}^{2} + r \hat{G}(u,v,r) ) \right\} \; .
        \label{origbfm14.2}
}

Introduce one last change of variables via
\leqn{origbfm15}{
\hat{v}_{k+1} = v_{k+1} + \frac{1}{2(k+1)}\pr{\tilde{k}} \left( f_{k+2}\right) \; .
}
Fix $X\in E_{+}$ and define $\hat{v} = \sum_{s\in \Ec} \hat{v}_{s+1}$.
Then using \eqref{origbfm10.2}, \eqref{origbfm13} and 
\eqref{origbfm15}, it can be shown that there exists a neighborhood
of $\mathcal{N}_{X}$ of $X$ in $E_{+}$, an 
$\epsilon > 0$,
and a sequence of analytic maps
\eqn{origbfm16}{
\mathcal{G}_{k} : \mathcal{N}_{X} \times E_{+} \times I_{\epsilon}(0) \times I_{\epsilon}(0)
\longrightarrow E^{\tilde{k}}_{0} \quad \forall \; k\in \Ec \; , 
}
such that
\leqn{origbfm17}{
r\hat{v}'_{k+1} = -2 (k+1) \hat{v}_{k+1} + r \mathcal{G}_{k}(u,\hat{v},\mu,r)
\quad \forall \; k\in \Ec \; .
} 
Also from  \eqref{origbfm6}, \eqref{origbfm14.2} and
\eqref{origbfm15}, it is not difficult to show that there exists
analytic maps
\eqn{origbfm18}{
\mathcal{H}_{k}  : E_{+} \times E_{+} \longrightarrow E^{\tilde{k}}_{+} \quad \forall \; k\in \Ec  
\AND
\mathcal{K}  : E_{+}\times E_{+} \times \mathbb{R} \times \mathbb{R} \longrightarrow
                \mathbb{R} \; ,
}
such that
\lalign{origbfm19}{
r u_{k+1}' & = r \mathcal{H}_{k}(u,\hat{v}) \quad \forall \; k\in  \Ec \; ,
\label{origbfm19.1} \\
r  \mu' & = -3\mu + r \mathcal{K}(u,\hat{v},\mu,r) \; . \label{origbfm19.2}
}

The system of differential equations \eqref{origbfm17}, 
\eqref{origbfm19.1} and  \eqref{origbfm19.2} are in the 
form for which  theorem \ref{BFM} applies. Applying this 
theorem shows that for fixed $X \in E_{+}$ there exist a 
unique solution 
$\{ u_{k+1}(r,Y), \hat{v}_{k+1}(r,Y), \mu(r,Y)\}$ to this
system of differential equations that is analytic in 
a neighborhood of $(r,Y) = (0,X)$ and that satisfies
\lalign{origbfm20}{
u_{s+1}(r,Y) &= Y_{s} + \text{O}(r)  \quad \forall\; s\in \Ec \;, 
\label{origbfm20.1} \\
\hat{v}_{s+1}(r,Y) &=  \text{O}(r)  \quad \forall\; s\in \Ec \;, \label{origbfm20.2} \\
\mu(r,Y) &= \text{O}(r) \nonumber \; , 
}
where $Y_{s} = \pr{\tilde{s}}\left(Y\right)$. From
equation \eqref{origbfm13}, it is then
clear that mass $m$ satisfies
\eqn{origbfm21}{
m(r)  = \text{O}(r^{3}) \; .
}
Also from \eqref{origbfm12}, \eqref{origbfm15}, \eqref{origbfm20.1} and \eqref{origbfm20.2} it
is not difficult to see that
\leqn{origbfm21a}{
P = \text{O}(r^4) \AND G = \text{O}(r^2) \; .
}

From the results of the previous section there exists
an orthonormal basis $\{\;\fb_{j}\;|\; j=1,2,\ldots M\;\}$ 
for $E_{+}$ consisting of eigenvectors for $A_{2}$, i.e,
$A_{2}(\fb_{j}) = k_{j}(k_{j}+1)\fb_{j}$. Thus we can introduce
new variables $\{\;\hat{u}_{j}(r)\; | \; j=1,2,\ldots M \;\}$ via
\leqn{origbfm22}{
\sum_{s\in \Ec} u_{s+1}(r) r^{s+1} = \sum_{j=1}^{M} \hat{u}_{j}(r) r^{k_{j}+1}
\fb_{j} \; .
}
From proposition \ref{hwv} we know that $M = |S_{\lambda}|$.
So we can write $S_{\lambda} = \{\;\alpha_{j}\; | \; j=1,2,\ldots M\}$ and
we get from proposition \ref{Eplus} that
$\{\; \mathbf{e}_{\alpha_{j}} \; | \; j = 1,2,\ldots M \}$
is also a basis for $E_{+}$. Therefore there exists a real non-singular
matrix $C_{ij}$ such that
\leqn{origbfm23}{
\fb_{j} = \sum_{k=1}^{M} C_{kj} \mathbf{e}_{\alpha_{k}} \; . 
}
Expand $\Op$ and $\Lp$ in the basis 
$\{\; \mathbf{e}_{\alpha_{j}} \; | \; j = 1,2,\ldots M \}$ as follows
\leqn{origbfm24}{
\text{$\Op = \sum_{j=1}^{M} w_{j,0} \mathbf{e}_{\alpha_{j}}$ and
$\Lp(r) = \sum_{j=1}^{M} w_{j}(r) \mathbf{e}_{\alpha_{j}}$ .}
}
Then results \eqref{origbfm1}, \eqref{origbfm22}, \eqref{origbfm23}, and
\eqref{origbfm24} imply that
\eqn{origbfm25}{
w_{i}(r) = w_{i,0} + \sum_{j=1}^{M} C_{ij}\hat{u}_{j}(r) r^{k_{j}+1} 
\quad i=1,2,\ldots M\;,
}
while from \eqref{origbfm20.1} and \eqref{origbfm22} it is clear that
\eqn{origbfm26}{
\hat{u}_{j}(r,Y) = \beta_{j}(Y) + \text{O}(r) \quad j=1,2\ldots,M\; ,
}
where $\beta_{j}(Y) = \rip{\fb_{j}}{Y}$.
\end{proof}

\begin{proof}[Proof of theorem \ref{existInf}]
The proof of this theorem involves finding a change of variables
to put the system of differential equations \eqref{feq1} and \eqref{feq3} into 
a form where theorem \ref{BFM} applies in a neighborhood of $z=0$ where
$z = \frac{1}{r}$. This proof is similar to the proof of
theorem \ref{exist0} with the exception that theorem \ref{Liebfmb}
is needed instead of theorem \ref{Liebfma}. 

Since $\Lp$ satisfies \eqref{realLp}, we can introduce new variables
$\{u_{s}(z) \; | \; s\in \Ec\;\}$ that satisfy

\leqn{inftybfm1}{
\Lp(z) = \Op  + \sum_{s\in \Ec} u_{s}(z)\, z^{s}
}
where $\Op = \Lp|_{z=0}$ and $u_{s}(z) \in E^{\tilde{s}}_{+}$ for all $z$ and $s \in \Ec$.
Because $E_{+} = \bigoplus_{q=1}^{I}  E^{q}_{+}$, it is obvious that this transformation
is invertible. Define
\eqn{inftybfm2}{
\chi_{s} = \left\{
\begin{array}{ll}
1 & \text{ if $s\in \Ec$} \\
0 & \text{ otherwise}
\end{array}
\right. \; .
}
Then we can write $\Lp(z) = \Op  + \sum_{k=0}^{\infty}\chi_{k} u_{k}(z)\,z^{k}$.
Substituting this in \eqref{vardefs5} shows that there exists an integer
$N_{1}$ such that
\eqn{inftybfm3}{
\Fc = -\sum_{k\in \Ec} A_{2}(u_{k})\,z^{k} + \sum_{k=1}^{N_{1}} f_{k} z^{k} \; .
}
where 
\alin{inftybfm4}{
f_{k} = & \half \sum_{j=1}^{k-1} \Bigl\{ [[\Op,c(\chi_{j} u_{j})] + 
        [\Om,\chi_{j} u_{j}], \chi_{k-j} u_{k-j}]   \nonumber \\
      &  + [[\chi_{j} u_{j},c(\chi_{k-j} u_{k-j})], \Op] 
        + \sum_{s=1}^{j-1}  [[\chi_{s} u_{s},c(\chi_{j-s} u_{j-s})], \chi_{k-j} u_{k-j} ]
        \Bigr\} \; .
}
But $A_{2}(u_{k}) = k(k+1) u_{k}$ for
every $k \in \Ec$ by lemma \ref{A2xp} and hence
\leqn{inftybfm5}{
\Fc = -\sum_{k\in \Ec} k(k+1) u_{k}\,z^{k} + \sum_{k=1}^{N_{1}} f_{k} z^{k}  \; .
}
Define
\leqn{inftybfm6}{
v_{s} = \dz{u}_{s} \quad \forall \; s \in \Ec \; .  
}
where $\dz{(\cdot)} = \frac{d\,}{dz}\,(\cdot)$.
Using \eqref{inftybfm1}, \eqref{inftybfm5} and \eqref{inftybfm6},
the EYM equation \eqref{feq3} can be written as
\lalign{inftybfm7}{
\sum_{k\in\Ec} \dz{v}_{k}\, z^{k+1} = & \sum_{k\in\Ec} -2(k+1)v_{k} z^{k} + 
\sum_{k\in\Ec} \left\{ \frac{2}{z}\left(1-\frac{1}{N}\right) v_{k}
+ \frac{1}{z^{2}} \left( \frac{1}{N}-1-2mz\right) k(k-1) u_{k}  \right. \nonumber \\
& \left. + \frac{4m}{z}\left(\frac{1}{N}-1\right) k u_{k} + \frac{2}{N}(2m-z^{2}P) v_{k}
-\frac{zP}{N} k u_{k} \right\}\, z^{k+1} + \sum_{k\in\Ec} 2m k(k+1) u_{k} z^{k} \nonumber\\
& -\frac{1}{N} \sum_{k=1}^{N_{1}} f_{k}\, z^{k-1} \label{inftybfm7.2}
}
Applying the projections $\pr{\tilde{k}}$ for every $ k \in \Ec$ to equation 
\eqref{inftybfm7.2} yields
\lalign{inftybfm8}{
z \dz{v}_{k} = &  -2(k+1)v_{k} + z\left\{ \frac{2}{z}\left(1-\frac{1}{N}\right) v_{k}
+ \frac{1}{z^{2}} \left( \frac{1}{N}-1-2mz\right) k(k-1) u_{k}  \right. \nonumber\\
& \left. + \frac{4m}{z}\left(\frac{1}{N}-1\right) k u_{k} + \frac{2}{N}(2m-z^{2}P) v_{k}
-\frac{zP}{N} k u_{k} \right\} +  2m k(k+1) u_{k} \nonumber \\
& -\frac{1}{z^{k}N} \sum_{s=0}^{N_{1}-1} \pr{\tilde{k}} \left( f_{s+1} \right) \, z^{s}  \quad
 \forall \; k\in \Ec \; .\label{inftybfm8.2} 
}
The last term in \eqref{inftybfm8.2} is the main obstruction to putting this
equation into a form where theorem \ref{BFM} applies.  It seems to contain
terms of order $z^{-s}$ $\:(s > 0)$ . But this is not the case as the results
of section \ref{BFM1} can be used to show that
\eqn{infybfm8a}{
\frac{1}{z^{k}N} \sum_{s=0}^{N_{1}-1} \pr{\tilde{k}}\left(f_{s+1}\right) \, z^{s}
= \frac{1}{N} \sum_{s=k}^{N_{1}-1} \pr{\tilde{k}}\left(f_{s+1}\right) \, z^{s-k} \; .
}
Namely, using proposition \ref{Eplus}, it can be shown that $f_{k} \in E_{+}$ for
all $k$. From the definition of the $u_{s}$ it is obvious that 
$\chi_{s} u_{s} \in \bigoplus^{\tilde{s}}_{q=1} E_{+}^{q}$ for
$1 \leq s \leq \ke_{I}$, and therefore by letting $Z_{0} = \Op$ and $Z_{k} = \chi_{k} u_{k}$ for
$k\geq 1$ we get  $f_{s+1} \in \bigoplus_{q=1}^{\tilde{s}} E_{+}^{q}$ via
theorem \ref{Liebfmb} . Consequently, 
for every $k\in \Ec$
\eqn{inftybfm9}{
\pr{\tilde{k}} \left( f_{s+1} \right) = 0 \quad \text{ if $s < k$} \; ,
}
because $k\in \Ec$ implies that $k = \ke_{\tilde{k}}$ and hence 
it follows for $s < k=\ke_{\tilde{k}}$ that $\tilde{s} < \tilde{k}$.
Therefore we can rewrite \eqref{inftybfm8.2} as
\lalign{inftybfm10}{
z \dz{v}_{k} = &  -2(k+1)v_{k} + z\left\{ \frac{2}{z}\left(1-\frac{1}{N}\right) v_{k}
+ \frac{1}{z^{2}} \left( \frac{1}{N}-1-3mz\right) k(k-1) u_{k}  \right. \nonumber\\
& + \frac{4m}{z}\left(\frac{1}{N}-1\right) k u_{k} + \frac{2}{N}(2m-z^{2}P) v_{k}
-\frac{zP}{N} k u_{k} - \frac{z}{N} \sum_{s=k+1}^{N_{1}-1} \pr{\tilde{k}} \left( f_{s+1} \right) 
\, z^{s-k-1} \nonumber \\
& \left. + \left(1-\frac{1}{N}\right) \pr{\tilde{k}} \left( f_{k+1} \right) \right\} 
+ 2m k(k+1) u_{k} - \pr{\tilde{k}} \left( f_{k+1} \right)  \quad
 \forall \; k\in \Ec \; .\label{inftybfm10.2} 
}
It is clear that there exists analytic functions
\eqn{inftybfm11}{
\hat{P}  : E_{+} \times \mathbb{R} \rightarrow \mathbb{R} \AND
\hat{G}  : E_{+} \times E_{+} \times \mathbb{R} \rightarrow \mathbb{R} \; ,
}
such that
\eqn{inftybfm12}{
P  = \hat{P}(u,z) \AND
G  = z^{4}\hat{G}(u,v,z) \; ,
}
where $ u = \sum_{s\in \Ec} u_{s} $ and $v = \sum_{s\in \Ec} v_{s}$.
The EYM equation \eqref{feq1} can then be written  as
\leqn{inftybfm13}{
z\dz{m} = z\left[ (2mz-1) z^{2} \hat{G}(u,v,z) - P(u,z) \right] \; .
}

Introduce one last change of variables via
\leqn{inftybfm14}{
\hat{v}_{k} = v_{k} + \frac{1}{2(k+1)}\pr{\tilde{k}} \left( f_{k+1}\right) - k m u_{k} \; .
}
Fix $ a > 0$ and define $\hat{v} = \sum_{s\in \Ec} \hat{v}_{s}$.
Then using \eqref{inftybfm10.2}, and \eqref{inftybfm14},
it can be shown that there exists and $\epsilon > 0$, 
and a sequence of analytic maps
\eqn{inftybfm15}{
\mathcal{G}_{k} : E_{+} \times E_{+} \times I_{\epsilon}(a) \times I_{\epsilon}(0)
\longrightarrow E^{\tilde{k}}_{0} \quad \forall \; k\in \Ec \; , 
}
such that
\leqn{inftybfm16}{
z\dz{\hat{v}}_{k} = -2(k+1) \hat{v}_{k} + z \mathcal{G}_{k}(u,\hat{v},m,z)
\quad \forall \; k\in \Ec \; .
} 
Also from  \eqref{inftybfm6}, \eqref{inftybfm13} and
\eqref{inftybfm14}, it is not hard to show that there exists
analytic maps
\leqn{inftybfm17}{
\mathcal{H}_{k}  : E_{+} \times E_{+}\times \mathbb{R} \longrightarrow E^{\tilde{k}}_{+} \quad 
\forall \; k\in \Ec  \AND
\mathcal{K}  : E_{+}\times E_{+} \times \mathbb{R} \times \mathbb{R} \longrightarrow
                \mathbb{R} \; ,
}
such that
\lalign{inftybfm18}{
z \dz{u}_{k} & = z \mathcal{H}_{k}(u,\hat{v},m) \quad \forall \; k\in  \Ec \; , \label{inftybfm18.1}\\
z  \dz{m} & = z \mathcal{K}(u,\hat{v},m,z) \; . \label{inftybfm18.2}
}
The system of differential equations \eqref{inftybfm16},
\eqref{inftybfm18.1} and  \eqref{inftybfm18.2} are in the
form for which  theorem \ref{BFM} applies. Applying this
theorem shows that for fixed $(X,a) \in E_{+} \times (0,\infty)$ there exist a
unique solution
$\{ u_{k}(z,Y,m_{\infty}), \hat{v}_{k}(z,Y,m_{\infty}), m(z,Y,m_{\infty})\}$ to
this system of differential equations that is analytic in 
a neighborhood of $(z,Y,m_{\infty}) = (0,X,a)$ and satisfies
\lalign{inftybfm19}{
u_{s}(z,Y,m_{\infty}) &= Y_{s} + \text{O}(z)  \quad \forall\; s\in \Ec \;,
\label{inftybfm19.1} \\
\hat{v}_{s}(z,Y,m_{\infty}) &=  \text{O}(z)  \quad \forall\; s\in \Ec \;,
\nonumber \\
m(z,Y,m_{\infty}) &= m_{\infty} + \text{O}(z) \nonumber \; ,
}
where $Y_{s} = \pr{\tilde{s}}\left(Y\right)$. 
Let  $\{\;\fb_{j}\;|\; j=1,2,\ldots M\;\}$ and
$\{\; \mathbf{e}_{\alpha_{j}} \; | \; j = 1,2,\ldots M \}$ be the same basis
for $E_{+}$ as introduced in the proof of theorem \ref{exist0}.
Then we can introduce new variables $\{\;\hat{u}_{j}(z)\; | \; j=1,2,\ldots M \;\}$ via
\leqn{inftybfm20}{
\sum_{s\in \Ec} u_{s}(z) z^{s} = \sum_{j=1}^{M} \hat{u}_{j}(z) z^{k_{j}}
\fb_{j} \; .
}
Expand $\Op$ and $\Lp$ in the basis 
$\{\; \mathbf{e}_{\alpha_{j}} \; | \; j = 1,2,\ldots M \}$ as follows
\leqn{inftybfm21}{
\text{$\Op = \sum_{j=1}^{M} w_{j,\infty} \mathbf{e}_{\alpha_{j}}$ and
$\Lp(z) = \sum_{j=1}^{M} w_{j}(z) \mathbf{e}_{\alpha_{j}}$ .}
}
Results \eqref{inftybfm1}, \eqref{inftybfm20}, \eqref{origbfm23}, and
\eqref{inftybfm21} then imply that
\eqn{inftybfm22}{
w_{i}(z) = w_{i,\infty} + \sum_{j=1}^{M} C_{ij}\hat{u}_{j}(z) z^{k_{j}}
\quad i=1,2,\ldots M\;,
}
while from \eqref{inftybfm19.1} and \eqref{inftybfm20} it is clear that
\eqn{inftybfm23}{
\hat{u}_{j}(z,Y,m_{\infty}) = \alpha_{j}(Y) + \text{O}(z) \quad j=1,2\ldots,M\; ,
}
where $\alpha_{j}(Y) = \rip{\fb_{j}}{Y}$.
\end{proof}

\begin{proof}[Proof of theorem \ref{existHor} ]
The proof of this theorem involves finding a change of variables
to put the system of differential equations \eqref{feq1} and \eqref{feq3} into
a form where theorem \ref{BFM} applies in a neighborhood of $r=r_{H}$.

Note that although we use the space $E_{+}$ which was
defined in section \ref{BFM1}, this proof does
not depend on the results of section \ref{BFM1}. Indeed,
$E_{+}$ can be replaced by $\sum_{\alpha \in S_{\lambda}} \mathbb{R}
\mathbf{e}_{\alpha}$ everywhere in the proof below and one
does not have to know that 
$E_{+} = \sum_{\alpha \in S_{\lambda}} \mathbb{R}\mathbf{e}_{\alpha}$,
which is the content of proposition \ref{Eplus}.
The notation $E_{+}$ is used for convenience.

Introduce new variables $t$, $\mu$, and $v$ via
\lalign{blackbfm1}{
t & = r-\rh \; , \nonumber \\
N & = t(\mu + \nu) \label{blackbfm1.2}\; , \\
v & = (\mu + \nu) \Lp' \; . \nonumber
}
where $\nu$ is a constant. 
Then
\leqn{blackbfm2}{
t\frac{d \Lp}{dt} = t\left(\frac{v}{\mu+\nu}\right) \; ,
}
and it is clear that there exists analytic maps
\eqn{blackbfm3}{
\hat{\Fc} : E_{+} \longrightarrow E_{+} \AND
\hat{P}  : E_{+} \longrightarrow \mathbb{R} 
}
such that
\eqn{blackbfm4}{
\hat{\Fc}(\Lp) = \Fc \AND \hat{P}(\Lp) = P \;. 
}
Assume $|\nu|>0$. Define a analytic map
\eqn{blackbfm5}{
\hat{G} : E_{+} \times I_{|\nu|}(0) \rightarrow \mathbb{R}
}
by
\eqn{blackbfm6}{
\hat{G}(X,a) = \frac{1}{2(a+\nu)^{2}}\norm{X}^{2} \; .  
}
Then
\eqn{blackbfm7}{
 G = \hat{G}(v,\mu) \; .
}
Using these new variables, we can write the EYM equations \eqref{feq1} and \eqref{feq3} as
\lalign{blackbfm8}{
t \frac{d\mu}{dt} = & -(\mu+\nu) + \frac{1}{\rh} - \frac{2}{\rh^{3}}\hat{P}(\Lp)         
+ t\left[ \frac{1}{t}\left( \frac{1}{t+\rh} - \frac{1}{\rh} \right) \right. \nonumber \\
& \left. -\frac{2}{t}\left( \frac{1}{(t+\rh)^{3}} - \frac{1}{\rh^{3}}\right) \hat{P}(\Lp)
+ \left(\frac{\mu+\nu}{t+\rh}\right)(1+2 \hat{G}(v,\mu))
\right] \; , \label{blackbfm8.2}
}
and
\leqn{blackbfm9}{
t\frac{dv}{dt} = -v - \frac{1}{(t+\rh)^{2}}\hat{\Fc}(\Lp) - 
t\left(\frac{2\hat{G}(v,\mu)}{t+\rh}\right)v \; ,
}
respectively. Introduce two new variables $\mu$ and $\hat{v}$ via
\lalign{blackbfm10}{
\hat{\mu} & = \mu + \nu - \frac{1}{\rh} + \frac{2}{\rh^{3}} \hat{P}(\Lp) \; , \label{blackbfm10.1}\\
\hat{v} & = v + \frac{1}{\rh^{2}} \hat{\Fc}(\Lp)  \; . \label{blackbfm10.2}
}
Define an analytic map
\eqn{blackbfm11}{
\gamma : E_{+} \times \mathbb{R} \longrightarrow \mathbb{R}
}
by
\eqn{blackbfm12}{
\gamma(X,a) = a-\nu + \frac{1}{\rh} - \frac{2}{\rh^{3}} \hat{P}(X) \; .
}
Fix a vector $Z \in E_{+}$ that satisfies
$ \|\frac{1}{\rh} - \frac{2}{\rh^{3}} \hat{P}(Z)\| > 0$. Then if we
set
\eqn{blackbfm13}{
\nu = \frac{1}{\rh} - \frac{2}{\rh^{3}} \hat{P}(Z) \; ,
}
we get $\gamma(Y,0) = 0$. So we can define an open neighborhood $D$ of
$(Z,0) \in E_{+}\times \mathbb{R}$ by
\eqn{blackbfm14}{
D = \{\;  (X,a) \; | \; \| \gamma(X,a) \| < \|\nu \| \; \} .
}
Then from \eqref{blackbfm2},  \eqref{blackbfm8.2}, \eqref{blackbfm9}, \eqref{blackbfm10.1}, and  
\eqref{blackbfm10.2}, it is not hard to show that there exists an $\epsilon > 0$ and
analytic maps
\alin{blackbfm15}{
\mathcal{G} &: E_{+} \times D  \longrightarrow \mathbb{R} \; , \\
\mathcal{H} &: E_{+} \times D \times I_{\epsilon}(0) \longrightarrow \mathbb{R} \; , \\
\mathcal{K} &:  E_{+} \times D \times I_{\epsilon}(0) \longrightarrow \mathbb{R} \; ,
}
such that
\lalign{blackbfm16}{
&t\frac{d \Lp}{dt}  = t\mathcal{G}(\hat{v},\Lp,\hat{\mu}) \; , \label{blackbfm16.1} \\
&t \frac{d \hat{v}}{dt}  = -\hat{v} + t \mathcal{H}(\hat{v},\Lp,\hat{\mu},t) \; , 
 \label{blackbfm16.2}\\
&t \frac{d \hat{\mu}}{dt} =  -\hat{\mu} + t\mathcal{K}(\hat{v},\Lp,\hat{\mu},t) \; .
 \label{blackbfm16.3}
}

The system of differential equations \eqref{blackbfm16.1},
\eqref{blackbfm16.2} and  \eqref{blackbfm16.3} is in the
form for which  theorem \ref{BFM} applies. Applying this
theorem shows that exists a unique solution
$\{ \Lp(t,Y), \hat{v}(t,Y), \hat{\mu}(t,Y)\}$ to this system of
differential equations that is analytic in
a neighborhood of $(t,Y) = (0,Z)$ and that satisfies
\lalign{blackbfm17}{
\Lp(t,Y) &= Z + \text{O}(t)  \;, \label{blackbfm17.1} \\
\hat{v}(t,Y) &=  \text{O}(t) \; ,\nonumber \\
\hat{\mu}(t,Y) &= \text{O}(t) \label{blackbfm17.3} \; .
}
Expand  $Z$ and $\Lp$ in the basis
$\{\; \mathbf{e}_{\alpha} \; | \; \alpha \in S_{\lambda} \}$ as follows
\eqn{balckbfm18}{
\text{ $Z = \sum_{\alpha \in  S_{\lambda} } w_{\alpha,\rh} \mathbf{e}_{\alpha}$ and
$\Lp(t) = \sum_{\alpha \in  S_{\lambda} } w_{\alpha}(t) \mathbf{e}_{\alpha}$ }.
}
Then equation \eqref{blackbfm17.1} shows that
\eqn{blackbfm19}{
 w_{\alpha}(t,Z) =  w_{\alpha,\rh} + \text{O}(t) \quad \forall \; \alpha \in S_{\lambda} \; .
}
It also not difficult to show that equations \eqref{blackbfm1.2},  \eqref{blackbfm10.1},  and \eqref{blackbfm17.3}
imply that
\eqn{blackbfm20}{
N(t,Z) = \nu t + \text{O}(t^{2}) \; .
} 
From this it follows immediately that
\eqn{lbackbfm21}{
N(\rh) = 0 \AND N'(\rh) = \nu \; .
}
\end{proof}

\end{document}